\documentclass[sigconf]{acmart}
\AtBeginDocument{%
  \providecommand\BibTeX{{%
    \normalfont B\kern-0.5em{\scshape i\kern-0.25em b}\kern-0.8em\TeX}}}

\usepackage{booktabs}
\usepackage{multirow}
\usepackage{enumitem}
\usepackage{titlesec}
\usepackage{caption}
\usepackage{subcaption}
\usepackage{float}

\def \pU {\Omega_u}
\def \pC {\Psi_c}
\copyrightyear{2022}
\acmYear{2022}
\setcopyright{rightsretained}
\acmConference[WWW '22]{Proceedings of the ACM Web Conference 2022}{April 25--29, 2022}{Virtual Event, Lyon, France}
\acmBooktitle{Proceedings of the ACM Web Conference 2022 (WWW '22), April 25--29, 2022, Virtual Event, Lyon, France}\acmDOI{10.1145/3485447.3512031}
\acmISBN{978-1-4503-9096-5/22/04}

\mathchardef\mhyphen="2D

\begin{document}

%%
%% The "title" command has an optional parameter,
%% allowing the author to define a "short title" to be used in page headers.
\title{Comparative Explanations of Recommendations}

%%
%% The "author" command and its associated commands are used to define
%% the authors and their affiliations.
%% Of note is the shared affiliation of the first two authors, and the
%% "authornote" and "authornotemark" commands
%% used to denote shared contribution to the research.
\author{Aobo Yang$^1$, Nan Wang$^1$, Renqin Cai$^1$, Hongbo Deng$^2$, Hongning Wang$^1$}

\affiliation
{
    $^{1}$University of Virginia, Charlottesville \country{USA}
}

\affiliation
{
    $^{2}$Alibaba Group, Hangzhou \country{China}
}

\email{{ay6gv, nw6a, rc7ne}@virginia.edu, dhb167148@alibaba-inc.com, hw5x@virginia.edu}

%%
%% By default, the full list of authors will be used in the page
%% headers. Often, this list is too long, and will overlap
%% other information printed in the page headers. This command allows
%% the author to define a more concise list
%% of authors' names for this purpose.
\renewcommand{\shortauthors}{Aobo Yang, Nan Wang, Renqin Cai, Hongbo Deng, Hongning Wang}

\newcommand{\model}{{CompExp}}

% \setlength{\abovedisplayskip}{2pt}
% \setlength{\belowdisplayskip}{2pt}
  
%%
%% The abstract is a short summary of the work to be presented in the
%% article.
\begin{abstract}
%In order for a user to adopt a system's recommendations, the user needs to first build trust in the system. Prior research shows that explanations help users make more informed decisions and increase their trust in the personalized systems. 
As recommendation is essentially a \emph{comparative (or ranking)} process, a good explanation should illustrate to users why an item is believed to be better than another, i.e., comparative explanations about the recommended items. 
Ideally, after reading the explanations, a user should reach the same ranking of items as the system's. Unfortunately, little research attention has yet been paid on such comparative explanations. 
% As recommendation is essentially a ranking problem, a good explanation should not only illustrate why the user should pay attention to the recommended items, but also why one item is believed to be better than another, i.e., \emph{comparative explanations}. 

In this work, we develop an extract-and-refine architecture to explain the relative comparisons among a set of ranked items from a recommender system. For each recommended item, we first extract one sentence from its associated reviews that best suits the desired comparison against a set of reference items. Then this extracted sentence is further articulated with respect to the target user through a generative model to better explain why the item is recommended. We design a new explanation quality metric based on BLEU to guide the end-to-end training of the extraction and refinement components, which avoids generation of generic content. Extensive offline evaluations on two large recommendation benchmark datasets and serious user studies against an array of state-of-the-art explainable recommendation algorithms demonstrate the necessity of comparative explanations and the effectiveness of our solution. 

% We develop an extract-and-refine framework to explain the ranking of items by a recommender system. Specifically, for a target recommended item, whose ranking is known to be better than another item, we first extract a sentence from target item's existing reviews to best support this comparison. Then we rewrite the extracted sentence to articulate the advantage. The extraction and refinement steps are trained in an end-to-end fashion to best optimize the quality of final explanations. Extensive offline evaluations on two large recommendation benchmark datasets and serious user study against an array of state-of-the-art explainable recommendation algorithms demonstrate the necessity of comparative explanations and the effectiveness of our solution. 
\end{abstract}

%%
%% The code below is generated by the tool at http://dl.acm.org/ccs.cfm.
%% Please copy and paste the code instead of the example below.

% \ccsdesc[500]{Information systems~Sentiment analysis}
% \ccsdesc[500]{Computing methodologies~Neural networks}
%%
%% Keywords. The author(s) should pick words that accurately describe
%% the work being presented. Separate the keywords with commas.

%%
%% This command processes the author and affiliation and title
%% information and builds the first part of the formatted document.
\maketitle

\section{Introduction}
Modern recommender systems fundamentally shape our everyday life
% , from the music we listen to, the latest news we read, to the restaurants and hotels we would visit 
\cite{koren2009matrix, he2017neural, sarwar2001item, rendle2010factorization, aggarwal2016recommender, wu2020deja, cai2021category}. 
As a result, how to explain the algorithm-made recommendations becomes crucial in building users' trust in the systems \cite{zhang2018explainable}. Previous research shows that explanations, which illustrate how the recommendations are generated \cite{ribeiro2016should, lundberg2017unified} or why the users should pay attention to the recommendations \cite{wang2018explainable, sun2020dual, yang2021explanation}, can notably strengthen user engagement with the system and better assist them in making informed decisions \cite{bilgic2005explaining, herlocker2000explaining, sinha2002role}. 

\begin{figure}[t]
    \centering
    \includegraphics[width=0.78\linewidth]{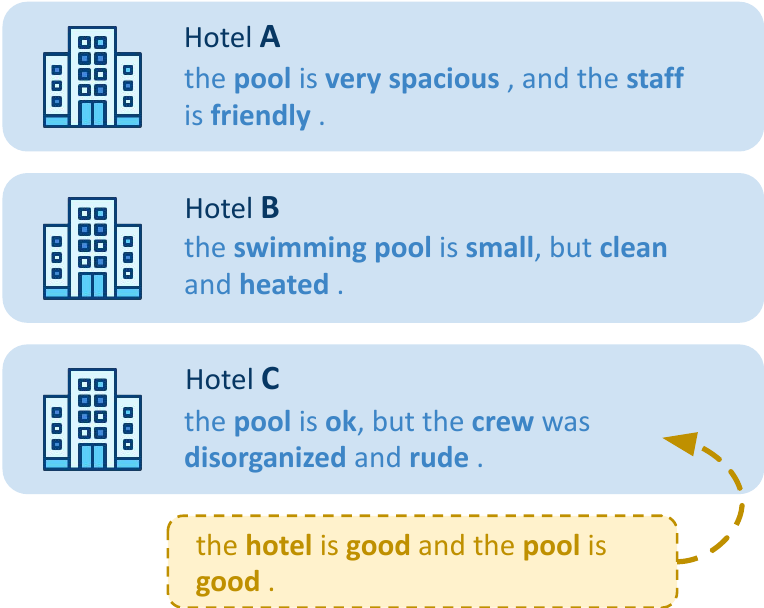}
    \caption{An illustration about the necessity of comparative explanations. The recommended Hotel A, B, C are listed in a descending order, with the provided explanations to justify the ranking. But if we replace Hotel C's explanation with the one in the dash box, users may no longer perceive the ranking of all three hotels.}
    \Description{Example of comparative explanations}
    \label{fig:example}
    \vspace{-3mm}
\end{figure}

When being presented with a list of recommendations, typically sorted in a descending order, a user needs to make a choice. In other words, the provided explanations should help users \emph{compare} the recommended items. 
Figure \ref{fig:example} illustrates the necessity of comparative explanations. 
%; and the fundamental bottleneck resides in the way they perform the learning of explanation generation. 
%More specifically, many mainstream explainable recommendation solutions leverage text generation techniques to synthesize natural language explanations  \cite{zhang2014explicit,wang2018explainable,wang2018reinforcement,tao2019the,chen2018neural,truong2019multimodal, sun2020dual}. 
%But due to the lack of dedicated corpus of textual explanations, most existing solutions (if not all) employed user-provided reviews as a proxy for explanation learning, where the text generation module is trained to reproduce the review content in a per-review or per-item basis. Such independent modeling of explanations cannot guarantee the generated content would align with the ranking of recommended items. %This design choice is based on the assumption that user reviews provide the rationale behind users' decisions, and once learnt the generated content will be related to how users make future decisions. 
%Recommendation is essentially a \emph{ranking problem} \cite{}. The recommendation algorithm needs to compare the items and rank with respect to align with users' preference \cite{}. Correspondingly, the explanations should also reflect the comparison among items and emphasize the advantage of the recommended item over others to illustrate the decision made by the algorithm. 
%To understand the necessity of comparative explanation in practice, take the hotel recommendations shown in Figure \ref{fig:example} for example. 
By reading the explanations for the hotels recommended in the figure, one can easily tell why the system ranks them in such an order. But if the system provided the explanation in the dashed box for Hotel C, it would confuse the users about the ranking, e.g., Hotel C becomes arguably comparable to top ranked Hotel A; but it was ranked at the bottom of the list. This unfortunately hurts users' trust in all three recommended hotels. 

Existing explainable recommendation solutions are not optimized to help users make such comparative decisions for two major reasons. First, the explanation of a recommended item is often independently generated without considering other items in the recommendation list. As shown in Figure \ref{fig:example}, one single low-quality generation (the one in the dashed box) might hamper a user's understanding over the entire list of recommendations. 
Second, the popularly adopted neural text generation techniques are known to be flawed of its generic content output \cite{holtzman2019curious, welleck2019neural}. Particularly, techniques like maximum likelihood training and sequence greedy decoding lead to short and repetitive sentences composed of globally frequent words \cite{weston2018retrieve}. Such generic content cannot fulfill the need to differentiate the recommended items. Consider the example shown in Figure \ref{fig:example} again, ``the hotel is good'' is a very generic explanation and thus not informative. Its vague description (e.g., the word ``good'') and lacks of specificity (e.g., the word ``hotel'') make it applicable to many hotels, such that users can hardly tell the relative comparison of the recommended items from such explanations.

In this work, we tackle the problem of comparative explanation generation to help users understand the comparisons between the recommended items. 
%Specifically, by pairing the  explanation with each review sentence of the user, the explanation is required to follow a set of comparative rankings. 
We focus on explaining how one item is compared with another; then by using a commonly shared set of items as references (e.g., items the user has reviewed before), the comparisons among the recommended items emerge. For example, if the explanations suggest item A is better than item B and item C is worse than item B, the comparison between A and C is apparent after reading the associated explanations.  
% Since there are already plenty of effective recommendation algorithms deployed in practice, we decide not to invent yet another. Instead, we assume the existence of a performing recommendation algorithm, and build our solution on top of its provided item rankings. 
Our solution is designed to generically work on top of other existing recommender systems.
We do not have any assumptions about how the recommendation algorithm ranks items (e.g., collaborative filtering \cite{sarwar2001item} or content-based \cite{balabanovic1997fab}), but only require it to provide a ranking score for each item to our model (i.e., ordinal ranking) which reflects a user's preference over the recommended item. This makes our solution readily applicable to explain plenty of effective recommendation algorithms deployed in practice.
%Our model learns the likelihood of a pair of explanations follows the order indicated by ratings and then leverage the learned knowledge to craft explanations that can best fit all the paired comparative rankings with the user's existing reviews. 

We design an extract-and-refine text generation architecture \cite{weston2018retrieve, guu2018generating} to explain the ranked items one at a time to the user, conditioned on their recommendation scores and associated reviews. We refer to the item to be explained in the ranked list as the target item, and user we are explaining to as the target user. 
First, the model extracts one sentence from the existing review sentences about the target item as a prototype, with a goal to maximize the likelihood of fitting the comparisons against the reviews written by the target user for other reference items. 
Then we refine the extracted prototype through a generative model to further polish the content for the target user. 
%Moreover, our hierarchical model design can alleviate the issue of generic content \cite{weston2018retrieve}. 
In this two stage procedure, the extraction module exploits the content already provided about the target item to ensure the relevance of generated explanations (e.g., avoid mentioning features that do not exist in the target item); and the refinement module further improve the explanation (e.g., informativeness and diversity of content) beyond the limitation of the existing content. We design a new explanation quality metric based on BLEU to guide the end-to-end training of the two modules, with a particular focus to penalize short and generic content in generated explanations.
%, but is limited to the available candidates which may be less optimal, e.g. unfamiliar wording or unaligned sentimental opinion \cite{yang2021explanation}. 
%Combining it with the refining neural text generator allow them to complement each other: extraction provides a specific perspective with details and refinement further polishes the wording and sentiment. 
%It also worth noting that our model does not embed another recommendation module and can be used to explain any recommendation model as long as it outputs a comparable recommendation score such as predicted ratings.

We compared the proposed solution with a rich set of state-of-the-art baselines for explanation generation on two large-scale recommendation datasets. 
% : RateBeer \cite{julian2012learning} and TripAdvisor \cite{wang2010latent}. 
Besides, we also conducted extensive user studies to have the generated explanations evaluated by real users. Positive results obtained on both offline and online experiments suggested the effectiveness of comparative explanations in assisting users to better understand the recommendations and make more informed choices.
 
\section{Related Work}
% \subsection{Explainable Recommendation}
Most explainable recommendation solutions exploit user reviews as the source of training data. They  either directly extract from reviews or synthesize content to mimic the reviews. 
Extraction-based solutions directly select representative text snippets from the target item's existing reviews. For example, NARRE \cite{chen2018neural} selects the most attentive reviews as the explanation, based on the attention that is originally learned to enrich the user and item representations for recommendation. CARP \cite{li2019capsule} uses the capsule network for the same purpose. \citet{wang2018reinforcement} adopt reinforcement learning to extract the most relevant review text that matches a given recommender system's rating prediction. \citet{xian2021ex3} extract attributes from reviews to explain a set of items based on users' preferences. However, as such solutions are restricted to an item's existing reviews, their effectiveness is subject to the availability and quality of existing content. For items with limited exposure, e.g., a new item, these solutions can hardly provide any informative explanations.  

Generation-based solutions synthesize textual explanations that are not limited to existing reviews. One branch focuses on predicting important aspects of an item (such as item features) from its associated reviews as explanations \cite{wang2018explainable, tao2019the, he2015trirank, ai2018learning, cai2017accounting}. For instance, MTER \cite{wang2018explainable} and FacT \cite{tao2019the} predict item features that are most important for a user to justify the recommendation. They rely on predefined text templates to deliver the predicted features. The other branch applies neural text generation techniques to synthesize natural language sentences. In particular, NRT \cite{li2017neural} models item recommendation and explanation generation in a shared user and item embedding space. It uses its predicted recommendation ratings as part of the initial state for explanation generation. MRG \cite{truong2019multimodal} integrates multiple modalities from user reviews, including ratings, text, and associated images, for multi-task explanation modeling. %Our solution differs from them as we emphasize the generation of \emph{comparative explanations} to justify the recommender's ranking prediction. 

Our work is closely related to two recent studies, DualPC \cite{sun2020dual} and SAER \cite{yang2021explanation}, which focus on strengthening the relation between recommendations and explanations.
Specifically, DualPC introduces duality regularization based on the joint probability of explanations and recommendations to improve the correlation between recommendations and generated explanations. SAER introduces the idea of sentiment alignment in explanation generation. However, both of them operate in a \emph{pointwise} fashion, i.e., independent explanation generation across items. Our solution focuses on explaining the comparisons between items. We should also emphasize our solution is to explain the comparison among a set of recommended items, rather than to find comparable items \cite{mcauley2015inferring,chen2020try}. 

There are also solutions exploiting other types of information for explainable recommendation, such as item-item relation \cite{chen2021temporal}, knowledge graph \cite{xian2019reinforcement} and social network \cite{ji2016jointly}. But they are clearly beyond the scope of this work. 
% a recommendation rating as pair-wise rankings with other items' ratings and hence models the explanation under the context of existing user reviews to emphasize the corresponding comparative relation of paired explanations.

% \subsection{Extract and Refine}
%Hierarchical models such as extract-and-refine \cite{} have been successfully applied in many natural language generation domains. NeuralEditor \cite{guu2018generating} is a generative language model that supports controllable editing to convert a sentence sampled from corpus to a new one. \citet{li2018delete} introduces several methods of mixing text operations of retrieve, delete and generate to transfer language style and sentiment. \citet{chen2018fast} implements separate extractor and abstractor to select sentences from article and then rewrites them into summaries. RetNRef \cite{weston2018retrieve} uses retrieve and refine in dialogue generation to avoid short and generic responses. Our solution is the first to apply the extract-and-refine workflow in the explanation generation problem. 
% Although we share the similar workflow, our model design differs from them greatly. 
%While most of existing work keep each the extractor and refiner independent, we connect and integrate them to enable end-to-end training. Moreover, our extractor does not only select prototypes for the downstream refiner but also provides it guidance about how to revise the prototypes.
\section{Comparative Explanation Generation}

Item recommendation in essence is a ranking problem: estimate a recommendation score for each item under a given user and rank the items accordingly, such that the utility of the recommendations can be maximized \cite{rendle2012bpr,karatzoglou2013learning}. 
Instead of explaining how the recommendation scores are obtained, our work emphasizes on explaining how the comparisons between the ranked items are derived. 
% objective of our work is to learn a generative explanation model that synthesizes textual explanations to explain how the comparisons between the ranked items are derived.

To learn the explanation model, we assume an existing corpus of item reviews from the intended application domain (e.g., hotel reviews). Each review is uniquely associated with a user $u$ and an item $c$, and a user-provided rating ${r}_c^u$ suggesting his/her opinion towards the item. 
% To simplify our discussion, we directly treat this opinion rating as the recommendation score ${r}_c^u$ (since both of them are input to our model, and most existing recommendation algorithms are estimated from such ratings \cite{rendle2012bpr,wang2018explainable}).
We group the reviews associated with user $u$ to construct his/her profile $\pU = \{(x^u_1, r^u_1), (x^u_2, r^u_2), ..., (x^u_m, r^u_m)\}$, where $x^u_i$ is the $i$-th review sentence extracted from user $u$'s reviews and $r^u_i$ is the corresponding opinion rating. $r^u_i$ can be easily obtained when the detailed aspect ratings are available \cite{wang2010latent}; otherwise off-the-shelf sentiment analysis methods can be used for the purpose (interested users can refer to \cite{wang2018explainable,zhang2014explicit} for more details). 
As regards cold-start for users without reviews, generic profiles can be used instead which sample reviews from similar users clustered by other non-review-related features, such as rating history. 
We create the item profile as $\pC = \{x^c_1, x^c_2, ..., x^c_n \}$, where $x^c_j$ is the $j$-th review sentence extracted from item $c$'s existing reviews. Unlike the user profile, the item profile does not include ratings. This is because the ratings from different users are not directly comparable, as individuals understand or use the numerical ratings differently. Our solution is agnostic to the number of entries in user profile $\pU$ and item profile $\pC$ in each user and item. 
% So it is more meaningful to compare the numerical ratings about different items from the same user.
%For example, with rating scale from one to five, three-star may mean "good" for a user while meaning "bad" for another. Therefore, we only utilize the ratings of the same user, i.e., $r^u_1, ..., r^u_m$.

%The explanation model is trained to characterize the observed comparisons in user reviews. For any tuple in a user's profile $\pU$, the model should maximize the probability of observing this tuple against all the other tuples in $\pU$. This can be considered as \emph{a set to sequence generation problem}. Consider a particular tuple $(x^u_i, r^u_i)$ from user $u$'s profile, 
%it can form a pairwise comparison with any other tuple $(x^u_i, r^u_i)$ in $U$. 
%the event that it forms a comparison to tuple $(x^u_j, r^u_j)$ can be described by the conditional probability $Q(x^u_i, r^u_i|x^u_j, r^u_j)$. Then the probability of observing tuple $(x^u_i, r^u_i)$ given the rest tuples in $\pU$ can be computed by summing the joint probability over all the other tuples $(x^u_j, r^u_j), \forall j \neq i$, i.e., $Q(x^u_i, r^u_i) = \sum_{j \neq i} Q(x^u_i, r^u_i|x^u_j, r^u_j)Q(x^u_j, r^u_j)$.
%Since we are only interested in the generation probability of the given text content $x^u_i$, instead of the entire tuple, we treat rating $r^u_i$ as given and model $Q(x^u_i|r^u_i, \pU)$ instead,

We impose a generative process for a tuple $(x, r^u_c)$ from user $u$ about item $c$ conditioned on $\pC$ and $\pU$. We assume when user $u$ is reviewing item $c$, he/she will first select an existing sentence from $\pC$ that is mostly related to the aspect he/she wants to cover about the item. Intuitively, this can be understood as the user will first browse existing reviews of the item to understand how the other users evaluated this item. Then he/she will rewrite this selected sentence to reflect his/her intended opinion and own writing style. This can be considered as \emph{a set to sequence generation problem}. For our purpose of explanation generation, we only concern the generation of opinionated text $x$. Hence, we take opinion rating $r^u_c$ as input, which leads us to the following formulation,
\begin{equation}
\label{eq_comp_U}
%    Q(x^u_i|r^u_i, \pU) = \sum_{j \neq i} Q(x^u_i|x^u_j, r^u_i, r^u_j)Q(x^u_j, r^u_j|r^u_i)
    P(x|u,c,r^u_c) = \sum_{x^c_j \in \pC} P_{ref}(x|x^c_j,r^u_c,\pU)P_{ext}(x^c_j|r^u_c,\pU)
\end{equation}
where $P_{ext}(x^c_j|r^u_c,\pU)$ specifies the probability that $x^c_j$ from item profile $\Psi_c$ will be selected by user $u$, and $P_{ref}(x|x^c_j,r^u_c,\pU)$ specifies the probability that user $u$ will rewrite $x^c_j$ into $x$. We name the resulting model Comparative Explainer, or \model{} in short.
%We still keep the prior as $Q(x^u_i, r^u_i)$ because the rating $r^u_j$ does not impact the probability of an existing tuple $(x^u_i, r^u_i)$. 
%$r^u_i$ and $r^u_j$ in the condition part indicate the desired comparison between the two items by user $u$, as simply suggested by their difference $\Delta^u_{ij} =r^u_i - r^u_j$. 

In Eq \eqref{eq_comp_U}, $P_{ext}(x^c_j|r^u_c,\pU)$ is essential to capture the comparative textual patterns embedded in user $u$'s historical opinionated text content. To understand this, we can simply rewrite its condition part: define $\Delta r^u_i=r^u_c-r^u_i$, we have $(r^u_c,\pU) = \{(x^u_i, \Delta r^u_i)\}^m_{i=1}$; hence, $P_{ext}(x^c_j|r^u_c,\pU)$ characterizes whether the sentence $x^c_j$ about item $c$ is qualified to characterize the desired opinion difference conditioned on user $u$'s historical content $\pU$ and target rating $r^u_c$. For example, a negative $\Delta r^u_i$ suggests the opinion conveyed in $x^c_j$ is expected to be less positive than that in $x^u_i$. On a similar note, $P_{ref}(x|x^c_j,r^u_c,\pU)$ quantifies if $x$ is a good rewriting of $x^c_j$ to satisfy the desired opinion rating $r^u_c$ for item $c$ by user $u$.

One can parameterize $P_{ext}(x^c_j|r^u_c,\pU)$ and $P_{ref}(x|x^c_j,r^u_c,\pU)$ and estimate the corresponding parameters based on the maximum likelihood principle over observations in $\pU$. However, data likelihood alone is insufficient to generate high-quality explanations, as we should also emphasize on fluency, brevity, and diversity of the generated explanations.  
To realize this generalized objective, assume a metric $\pi(x|u,c)$ that measures the quality of generated explanation $x$ for user $u$ about item $c$, the training objective of \model{} is set to maximize the expected quality of its generated explanations under $\pi(x|u,c)$,
\begin{equation}
\label{objective}
    J = \mathbb{E}_{x \sim P(x|u,c,r^u_c)}[\pi(x|u,c)]
\end{equation}
In this work, we present a customized BLUE score specifically for the comparative explanation generation problem to penalize short and generic content. 

Next, we dive into the detailed design of \model{} in Section \ref{sec_arch}, then present our metric $\pi(x|u,c)$ for parameter estimation in Section \ref{sec_idfBLEU} and \ref{sec_reward}, and finally illustrate how to estimate each component in \model{} end-to-end in Section \ref{sec_train}.

\subsection{Extract-and-Refine Architecture}
\label{sec_arch}

\begin{figure}[h]
    \centering
    \includegraphics[width=0.925\linewidth]{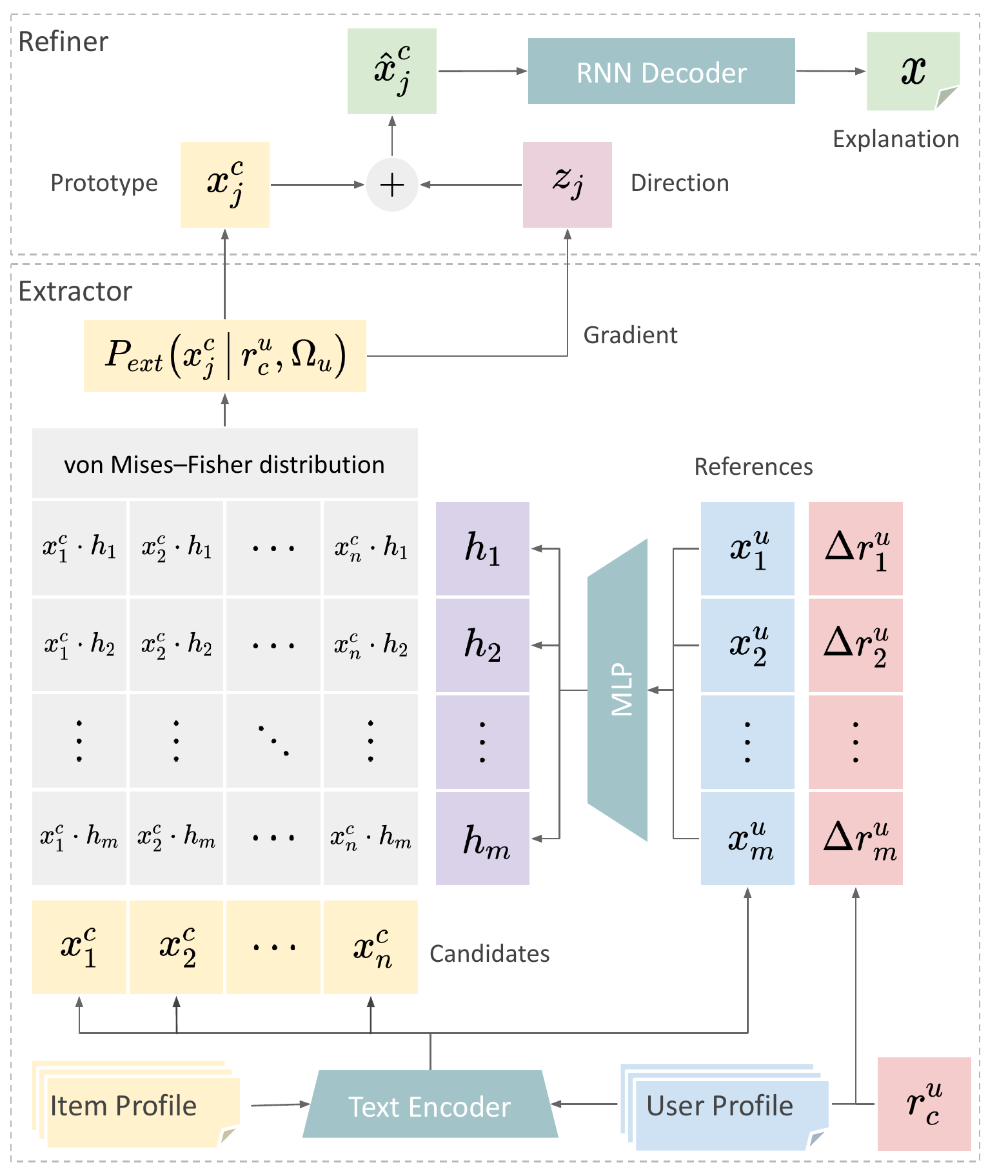}
    \vspace{-2mm}
    \caption{The extract-and-refine model architecture for \model{}. The extractor  extracts a candidate sentence from item $c$'s profile as a prototype for explanation generation; and the refiner rewrites this sentence to optimize the desired quality metric for comparative explanation.}
    \label{fig:model}
    \vspace{-2mm}
\end{figure}

Our proposed model architecture for \model{} is shown in Figure \ref{fig:model}, which in a nutshell is a fully connected hierarchical neural network. The explanations for a user item pair $(u,c)$ is generated via an extract-and-refine process, formally described in Eq \eqref{eq_comp_U}. 
%First, \model{} retrieves a sentence $x^c_j$ from the item $c$'s profile $\pC$ as a prototype; and then it rewrites the prototype to generate the final explanation $x$, with a goal to better optimize the desired quality metric $\pi(x|u,c)$. 
%This design is based on our observation that many explanations of the same item are very similar in lexicon and can be transformed to each other with minor tweaks. This is understandable since although the explanations may contain users' subjective opinions and personalized language style, their main topics are unavoidably scoped by the exact same objective facts of the item. For example, if a hotel is located in the beach, many existing explanations will highlight such feature and it is unlikely that any explanations will claim the location otherwise, say downtown. 
Comparing to existing pure generation-based explanation methods \cite{li2017neural,sun2020dual,yang2021explanation}, one added benefit of our solution is to ensure faithfulness of the generated explanations: it avoids mentioning attributes that are not relevant to the target item. 
To address the limitations in directly using existing content, e.g., unaligned content style or sentiment polarity, the refinement step further rewrites the extracted sentence to make its content better fit for the purpose of comparative explanation, e.g., improve the quality defined by $\pi(x|u,c)$. 

%explanations can serve as a good reference for future explanations. Moreover, the extracted prototype can also help reduce the untruthful "model made-up" content, e.g. mentioning "pool" for a hotel without pool, because the generative model is now conditioned upon the human-written prototype. 

%The generative explanation process $P(x|u,c)$ can be specified by the marginal probability over all possible prototypes:
%\begin{equation}
%\label{extref}
%    P(x|u,c) = \sum_{x^c_i \in \pC} P_{ref}(x|x^c_i,\pU)P_{ext}(x^c_i|\pU)
%\end{equation}
%where $P_{ext}(x^c_i|\pU)$ is the probability of extracting sentence $x^c_i$ from item $c$'s profile for user $u$, and $P_{ref}(x|x^c_i,\pU)$ is the probability of rewriting  $x^c_i$ into $x$. 
%We use policy gradient to optimize Eq \eqref{objective} under the generative process defined by Eq \eqref{extref},
%\begin{equation*}
%    \nabla_\Theta J \approx \pi(x|u,c) \nabla_\Theta \log P_{ref}(x|x^c_i,\pU) + \pi(x|u,c) \nabla_\Theta \log P_{ext}(x^c_i|\pU)
%\end{equation*}
%where $\Theta$ stands for all model parameters in \model{}. 
We refer to $P_{ext}(x^c_j|r^u_c,\pU)$ as the extractor and $P_{ref}(x|x^c_j,r^u_c,\pU)$ as the refiner. Next, we will zoom into each component to discuss its design principle and technical details.

\subsubsection{Extractor.} 
% \noindent\textbf{$\bullet$ Extractor.} 
The extractor's goal is to select a prototype sentence $x^c_j$ from item $c$'s profile $\pC$ for a given opinion rating ${r}^u_c$ that best satisfies the comparativeness suggested by the user profile $\pU$. 
We refer to $x^c_j\in\pC$ as an extraction candidate and $x^u_i\in \pU$ as a reference. The extractor adopts a bidirectional GRU \cite{chung2014empirical} as the universal text encoder to convert the extraction candidates and references into continuous embedding vectors. Since the pairwise comparison specified by $\Delta r^u_i$ is a scalar, we use a one-hot vector to encode it when the ratings are discrete, otherwise we use a non-linear multi-layer perceptron (MLP) as the rating encoder. 

% Inspired by the linear algebra operations enabled by the learnt word embeddings, we would like to learn a set of text embeddings that transform the reference sentence in the latent text space to its ideal comparative explanation vector based on the desired rating difference, written as $f(x^u_j,\Delta r^u_j) \to h_j $, where $h_j$ is the ideal text vector for the comparative explanation. Intuitively, in the one dimensional rating space, we can easily recover the target item's rating $\hat{r}^u_c$ from the rating of the reference item ${r}^u_j$ and rating difference $\Delta r^u_j$. Then, as an analogy, we also want to achieve the same in the learnt text embedding space to recover the ideal comparative explanation with respect to a reference sentence and desired rating difference.

Intuitively, in the one dimensional rating space, we can easily recover the intended sentence's rating ${r}^u_c$ from the rating of the reference sentence ${r}^u_i$ and required rating difference $\Delta r^u_i$. As an analogy, we consider the rating difference vector as the transform direction that suggests the ideal comparative explanation in the latent text space from a reference sentence $x^u_i$, denoted as $f(x^u_i,\Delta r^u_i) \to h_i$. As a result, $h_i$ is the text embedding vector for the ideal comparative explanation. The extractor implements such a transformation using an MLP taking the concatenation of the text embedding and rating difference embedding vectors as input. 

Given the desired comparative explanation $h_i$, the extraction candidates can be evaluated by their similarities towards $h_i$. This specifies a directional distribution $Q(x;h_i)$ centered on $h_i$ in the latent text embedding space. Since cosine is a commonly used similarity metric for text embeddings, we formulate $Q(x;h_i)$ as  a von Mises-Fisher distribution \cite{guu2018generating} over all the extraction candidates,
\begin{equation*}
    Q(x;h_i) \propto f_{vMF}(x; h_i, \kappa) = C_p(\kappa) e^{\kappa \cos(x, h_i)}
\end{equation*}
where $f_{vMF}(\cdot)$ is the probability density function, $\kappa$ is the concentration parameter, and $C_p(\kappa)$ is a normalization function about $k$. 
Because each reference sentence $x^u_i$ will suggest a different directional distribution, we extend the von Mises-Fisher distribution to cover multiple centriods and define $P_{ext}(x^c_j|r^u_c,\pU)$ as follows,
\begin{equation}
\label{eq_extract_prob}
P_{ext}(x^c_j|r^u_c,\pU)\propto \sum_{x^u_i\in\pU} f_{vMF}\Big(x^c_j; f(x^u_i, \Delta r^u_i), \kappa\Big)
\end{equation}
Intuitively, in Eq \eqref{eq_extract_prob}, each ideal embedding $h_i$ suggests which extraction candidate better fits the comparativeness embedded in $\pU$. The summation over $\pU$ aggregates each reference sentence's evaluation on candidate sentence $x^c_j$. $\kappa$ is kept as a hyper-parameter which shapes the extraction probability distribution: a larger $\kappa$ value leads to a skewer distribution. We can use it to control the exploration of the extraction candidates during the policy gradient based model training, which will be introduced in Section \ref{sec_train}.

% \noindent\textbf{$\bullet$ Refiner.} 
\subsubsection{Refiner.} 
The objective of the refiner is to rewrite the extracted prototype to further improve the quality metric $\pi(x|u,c)$. As we argued before, a better explanation should be more supportive to the pairwise comparison required by the user profile. Therefore, assuming the refiner successfully turns the prototype $x^c_j$ into a better framed sentence $\hat{x}^c_j$ about the item $c$ for user $u$, then when we give $\hat{x}^c_j$ back to the extractor together with $x^c_j$, the extractor should prefer the revised version over the original one. Otherwise, we should keep refining $\hat{x}^c_j$ until the extractor believes it can no longer be improved. Hence, the refiner needs to find a direction such that $P_{ext}(x^c_j|r^u_c, \pU) < P_{ext}(\hat{x}^c_j|r^u_c, \pU)$, which is exactly suggested by the gradient of $P_{ext}(x^c_j|r^u_c, \pU)$ with respect to $x^c_j$, i.e., the fastest direction for $x^c_j$ to increase the value of $P_{ext}(x^c_j|r^u_c, \pU)$. As a result, our refiner simply pushes the text embedding vector of $x^c_j$ alone this gradient direction: 
%Specifically, it directly reuses the prototype's text embedding vector from the extractor's text encoder and calculates the gradient of $Q(x^c_i|\hat{r}^u_c, \pU)$ with respect to $x^c_i$ and normalize it to a unit vector as the refining direction vector $z_i$,
\begin{equation*}
\begin{split}
    {z}_j &= \nabla_{x^c_j} P_{ext}(x^c_j|r^u_c, \pU) \\
     & \propto  \sum^m_i e^{\kappa \cos(x^c_j, h_i)} \big[ \frac{h_i}{|x^c_j||h_i|} - \cos(x^c_j, h_i) \frac{x^c_j}{|x^c_j|^2} \big] %  \\
    %z_i &= \hat{z}_i / |\hat{z}_i|
\end{split}
\end{equation*}
Since the refinement step should only polish the extracted prototype instead of dramatically changing it, 
% we constrain the step size in this refinement. Empirically, 
we normalize the gradient to a unit vector and restrict the step size to one in all cases, i.e., $\hat{x}^c_j = x^c_j + z_j / |z_j|$. 
At last, we include a single-layer GRU with attention \cite{luong2015effective} as the text decoder to convert the refined text vector $\hat{x}^c_j$ to the final explanation sentence $x$. 

Connecting these two modules together, \model{} generates explanations for a ranked list of recommended items one at a time. To understand why the generated explanations carry comparativeness, we can consider the user's profile $\pU$ as an anchor. Because all the explanations are generated against this anchor, the comparisons among the explanations emerge. 

\subsection{Explanation Quality Metric}
\label{sec_idfBLEU}
To train \model{} under Eq \eqref{objective}, we need to define the explanation quality metric $\pi(x|u,c)$. There is no commonly agreed offline metric for explanation quality in the community yet. And obtaining real user feedback is not feasible for offline model training. 
Currently, most of explainable recommendation solutions \cite{sun2020dual, li2017neural, yang2021explanation} adopt metrics measuring the overlapping content between the generated explanations and user reviews, such as BLEU \cite{papineni2002bleu}.
%, which is a popularly used metric to evaluate the generated content's closeness to one or more references. 
%Since the user reviews are intrinsically comparative with respect to the recommendation ranking, the explanations should be close to the sentences of the review. 

However, the BLEU metric, which is initially designed for machine translation, is problematic in explanation evaluation for at least two important reasons. 
First, it is biased towards shorter sentences. As a precision-based metric, BLEU overcomes the short-length issue by introducing the brevity penalty, which down-scales the precision when the generated length is smaller than its ``best match length'' \cite{papineni2002bleu}. The ``best match length'' design is reasonable in machine translation, because all reference sentences are valid translations covering the information contained in the source language, regardless of their length differences. However, when using review sentences as proxies of explanations, the reference sentences from one review can describe totally different aspects of the same item and vary significantly in length and information contained. Since short-length generation benefits precision (less prone to erroneous word choices), BLEU favors explanations exploiting the short references as the ``best match''. As a result, it pushes the models to generate explanations that are generally much shorter than the average sentence length in a review, and hence fails to explain the item in details. 
Second, though precision-based, BLEU is incapable to differentiate the importance of different words in a reference sentence. Words are valued equally in machine translation, but their impact in explanations varies significantly to users. 
%Especially in comparative explanations, we emphasize more on informative content to differentiate different items. 
For example, in Figure \ref{fig:example}, the feature and descriptive words such as ``swimming pool'' and ``friendly'' help users better understand the target item than a very frequent but generic word, like ``hotel'' and ``good''. %Some solutions \cite{wang2018explainable, tao2019the} even directly list such critical feature words as explanations. 
BLEU's indiscrimination to words unavoidably favors the explanations with more generic content due to their higher chance of appearance. We later demonstrate how the BLEU metric led to both short and generic explanations in our experiments.

To design a more appropriate metric to evaluate the explanation quality and better guide our model training, we propose IDF-BLEU, i.e., Inverse Document Frequency (IDF) enhanced BLEU. It introduces three changes on top of BLEU to balance the important factors in explanations: length, content overlapping, and content rarity. 

First, to penalize an overly short generation, we replace the ``best match length'' in the brevity factor with the average length of sentences from all reviews,
\begin{equation*}
BP_{len} = e^{\min(1 - \frac{l_r}{l_x}, 0)}
\end{equation*}
where $l_r$ and $l_x$ is the average length of references and the length of the explanation respectively. Second, to differentiate the importance of different words, we introduce IDF to measure the value of n-grams and use it to reweigh the precision in BLEU. We compute the IDF of word $g$ by the number of sentences where it occurs,
\begin{equation*}
    IDF(g) =  log \frac{S}{s_g} + 1
\end{equation*}
where $S$ is the total number of review sentences in the training corpus and $s_g$ is the number of sentences containing word $g$. We approximate the IDF of an n-gram by the largest IDF of its constituent words. Then the clipped n-gram precision in BLEU is modified as
\begin{equation}
\label{clip_precision}
    p_n = \frac{ \sum_{g^n \in x} IDF(g^n) \cdot Count_{clip}(g^n) }{ \sum_{g^n \in x} IDF(g^n) \cdot Count(g^n) }
\end{equation}
where $g^n$ represents the n-gram and $Count_{clip}(g^n)$ is the BLEU's operation to calculate the count of $g^n$ in sentence $x$ while being clipped by the corresponding maximum count in the references. Through the reweighing, correctly predicting an informative word becomes more rewarding than a generic word. However, it alone cannot evaluate content rarity, since the precision-based metric cannot punish sentences for not including rare words. Therefore, at last, inspired by the length brevity factor in original BLEU, we introduce a similar IDF brevity factor to punish sentences lacking words with high IDF,
\begin{equation*}
BP_{IDF} = e^{\min(1 - \frac{d_r}{d_x}, 0)}
\end{equation*}
where $d_x$ is the average IDF per word $d_x = \sum_{g \in x} IDF(g) / l_x $ and $d_r$ is corresponding average value in references. Then combining them forms our IDF-BLEU,
\begin{equation}
\label{eq_BLEU}
    IDF \mhyphen BLEU = BP_{len} \cdot BP_{IDF} \cdot \exp \Big( \sum^N_{n=1} w_n \log p_n \Big)
\end{equation}
where $w_n$ is BLEU's parameter used as the weight of the n-gram precision. We use the proposed IDF-BLEU as the quality metric $\pi(x|u,c)$ for \model{} training.

\subsection{Hierarchical Rewards}
\label{sec_reward}
\model{} is a fully connected neural network which can be trained end-to-end with the gradient derived from Eq \eqref{objective}. 
However, blind end-to-end training faces the risk that the model violates the purpose of our designed extract-and-refine procedure, as the model has a great degree of freedom to arbitrarily push the prototype $x^c_j$ in the continuous vector space to optimize Eq \eqref{objective}. For example, it could disregard the extracted prototype and generate totally irrelevant content to the target item $c$ in the refiner. 
%We cannot guarantee the extractor uses it to represent a potential explanation candidate or the refiner respects the content it is supposed to represent. 

To enforce the extract-and-refine workflow, we introduce additional intrinsic reward \cite{vezhnevets2017feudal} for each layer respectively to regularize their behaviours. 
%by encouraging the extractor to select prototypes which act as good explanations themselves and rewarding refiner for accurately retain the prototype. 
Specifically, as IDF-BLEU is used to measure the explanation quality in Eq \eqref{objective}, we directly use the extracted sentence's IDF-BLEU to reward the extractor, i.e., introduce $\pi_{ext}(x^c_j|u,c) = IDF \mhyphen BLEU(x^c_j)$. 
For the refiner, we discourage it in pushing the final generation too far away from the extracted one. Inspired by the clipped precision in Eq \eqref{clip_precision}, we propose a clipped recall to measure how many words from the selected sentence $x^c_j$ are still covered in the refined sentence,
\begin{equation}
\label{eq_clipped_recall}
    a_n = \frac{ \sum_{g^n \in x^c_j} IDF(g^n) \cdot min[ Count_{clip}(g^n) , Count_{x}(g^n) ] }{ \sum_{g^n \in x^c_j} IDF(g^n) \cdot Count_{clip}(g^n) }
\end{equation}
where $Count_{clip}(g^n)$ is the clipped count of n-gram $g^n$ towards the references like in BLEU, and $Count_{x}(g^n)$ is the count of $g^n$ in the refined explanation $x$. In other words, the denominator is the prototype's overlap with the target references and the numerator is the overlap among the prototype, references, and the final explanation.
% Compared with classical recall definition, the clipped recall only encourage the refiner to keep the n-grams that are actually presented in the references, instead of retaining the entire prototype.
We did not use classical recall definition because it would reward the refiner to retain the entire prototype.
We only encourage the refiner to keep the n-grams that are actually presented in the references.
%The additional clipping by the references is how we define the "correct" content in the prototype. It only encourages the refiner to stick with prototype' successful overlap with the references and does not punish trying to change the other unmatched content. 
We compute the refiner's intrinsic reward by aggregating the clipped recall over different n-grams $\pi_{ref}(x, x^c_j) = \exp \big( \sum^N_{n=1} w_n \log a_n \big)$. We did not provide this reward to the extractor, because it biases the extractor to short and generic candidates which are easier for the refiner to cover.

With the hierarchical intrinsic rewards introduced for each component, we can optimize Eq \eqref{objective} by policy gradient as
\begin{equation*}
\begin{split}
    \nabla_\Theta J \approx & [\lambda_1 \pi(x|u,c) + \lambda_2 \pi_{ref}(x, x^c_j)] \nabla_\Theta \log P_{ref}(x|x^c_j,r^u_c,\pU) \\
    &+ [\lambda_3 \pi(x|u,c) + \lambda_4 \pi_{ext}(x^c_j)] \nabla_\Theta \log P_{ext}(x^c_j|r^u_c,\pU)
\end{split}
\end{equation*}
where $\lambda_1$ to $\lambda_4$ are coefficients to adjust the importance of each reward, and $\Theta$ stands for the model parameters in \model{}.

\subsection{Model Training}
\label{sec_train}
The whole model training process can be organized into two steps: pre-training and fine-tuning. The pre-training step aims to bootstrap the extractor and refiner independently. To prepare the extractor to recognize the comparative relationships among sentences, we treat every observed review sentence as the extraction target and train the extractor to maximize its negative log-likelihood with regard to the corresponding user and item profiles. 

It is important to pre-train the refiner as a generative language model, because it would be very inefficient to learn all the natural language model parameters only through the end-to-end training. However, we do not have any paired sentences to pre-train the refiner. We borrowed the method introduced in \cite{guu2018generating, weston2018retrieve} to manually craft such pairs. Specifically, for every sentence, we compute its cosine similarity against all other sentences in the same item profile in the latent embedding space, and select the most similar one to pair with. Then we use this dataset to pre-train the refiner
% : rely on the extractor to encode the prototype and calculate the direction vector; feed them into the refiner and supervise the decoding of the ground-truth explanation 
with negative log-likelihood loss. 

In the fine-tuning stage, we concatenate the pre-trained layers and conduct the end-to-end training with policy gradient. 
%Unlike in pre-training, the ``ground-truth'' explanations are no longer included in the extraction candidates to force the model learn to utilize available candidates, as in actual testing phase, there is no such ``ground-truth'' explanations to choose from. 
To make the policy gradient training more resilient to variance and converge faster, it is important to have a baseline to update the model with reward advantages instead of using the rewards directly. We apply Monte Carlo sampling in both extractor and refiner to have multiple explanations, and use their mean rewards as the baseline.
\renewcommand{\arraystretch}{0.9}
\section{Experimental Evaluations}

%In this section, 
%we show that IDF-BLEU is a more reasonable metric than BLEU to evaluate the explanations. 
We demonstrate empirically that \model{} can generate improved explanations compared to state-of-the-art explainable recommendation algorithms. We conduct experiments on two different recommendation scenarios: RateBeer reviews with single-ratings \cite{julian2012learning} and TripAdvisor reviews with \emph{multi-aspect ratings} \cite{wang2010latent}.

% Moreover, they test if our model's explanations do capture the comparative ranking. Our experiments are conducted on two different recommendation scenarios: traditional single-rating with Ratebeer reviews dataset \cite{julian2012learning} and multi-aspects ratings with TripAdvisor dataset \cite{wang2010latent}. Our model is compared against a set of state-of-the-art baselines to prove the improvement from our model.

\subsection{Experiment Setup}

As our solution only focuses on explanation generation, it can be applied to any recommendation algorithm of choice. In our experiments, we directly use the ground-truth review ratings as the recommendation score to factor out any deviation or noise introduced by specific recommendation algorithms. For completeness, we also empirically studied the impact from input ratings if switched to a real recommendation algorithm's predictions. 

\begin{table}[t]
    \caption{Summary of the processed datasets.}
    \vspace{-2mm}
    \label{tab:stats}
    \begin{tabular}{|c|cccc|}
    \hline
    Dataset  & \# Users & \# Items & \# Reviews & \ Rating Range \\ 
    \hline
    RateBeer & 6,566    & 19,876    & 2,236,278  & 0 - 20           \\ 
    TripAdvisor     & 4,954   & 4,493   & 287,879  & 1 - 5           \\ 
    \hline
    \end{tabular}
    \vspace{-4mm}
\end{table}

\subsubsection{Data Pre-Processing}
In the RateBeer dataset, we segment each review into sentences, and label them with the overall ratings from their original reviews. In the TripAdvisor dataset, there are separate ratings for five aspects including \emph{service, room, location, value and cleanliness}. Therefore, each TripAdvisor review is expected to be a mix of a user's opinions on these different aspects about the item. We segment sentences in a TripAdvisor review to different aspects using the boot-strapping method from \cite{wang2010latent} and assign resulting sentences the corresponding aspect ratings. 
% Sentences that cannot be linked to any aspect are abandoned. 
These two datasets evaluate \model{} under different scenarios: overall opinion vs., aspect-specific opinion. We also adopt the recursive filtering \cite{wang2018explainable} to alleviate the data sparsity. The statistics of the processed datasets are summarized in Table \ref{tab:stats}.

% Although reviews are good source of explanations, not all of them are informative to users. Many previous work \cite{chen2018neural, wang2018reinforcement, sun2020dual} directly assumes the whole review as an explanation, but a recent work \cite{ni2019justifying} suggests that a large portion of review sentences are only about subjective experience which provides no meaningful information for other users to understand the recommendation as for the purpose of explanation, e.g., ``\textit{we spent two nights in this hotel.}''. Therefore, we should filter out such meaningless sentences from the review datasets. For the Ratebeer dataset, we use the method introduced in \cite{yang2021explanation} to keep only sentences that describe certain features of items as explanations. For the TripAdvisor, since review sentences have already been mapped to aspects, they can be treated as aspect descriptions and hence can be kept for explanations.

% We also adopt the recursive filtering \cite{wang2018explainable} to alleviate the common sparsity issue in the review datasets.
% The statistics of the processed datasets are summarized in Table \ref{tab:stats}. We build the vocabulary of each dataset by selecting the 20,000 most frequent words and mapping others to unknown. Each dataset is split into 70\% for training, 15\% for validation, and 15\% for testing.

\subsubsection{Baselines}

We compared with three explainable recommendation baselines that generate natural language explanations, covering both extraction-based and generation-based solutions. 
\begin{itemize}[leftmargin=*]
  \item[-] \textbf{NARRE}: Neural Attentional Regression model with Review-level Explanations \cite{chen2018neural}. It is an extraction-based solution. It learns the usefulness of the existing reviews through attention and selects the most attentive reviews as the explanation.
  \item[-] \textbf{NRT}: Neural Rating and Tips Generation \cite{li2017neural}. It is a generation-based solution. It models rating regression and content generation as a multi-task learning problem with shared latent space. Content is generated from its neural language model component. %It strengthens the correlation between rating and content by introducing the predicted ratings into the content generation. It is originally proposed for tip generation, but can be seamlessly adapted to generating explanations.
%   since tips play a similar role as explanations in recommendations.
  \item[-] \textbf{SAER}: Sentiment Aligned Explainable Recommendation \cite{yang2021explanation}. This is another generation-based solution using multi-task learning to model rating regression and explanation generation. But it focuses specifically on the sentiment alignment between the predicted rating and generated explanation. %It implements a sentiment regularizer and a constrained decoding method to enforce the sentiment in the explanation to defend the predicted rating.
%   in both training and inference phases.
\end{itemize}

We include three variants of \model{} to better demonstrate the effect of each component in it:
\begin{itemize}[leftmargin=*]
    \item[-] \textbf{\model{}-Ext}: the extractor of our solution. It directly uses the selected sentences as explanations without any refinement. This variant helps us study how the extractor works and also serves as a fair counterpart for the other extraction-based baseline. 
    \item[-] \textbf{\model{}-Pretrain}: our model with pre-training only, which is a simple concatenation of the separately trained extractor and refiner without joint training. We compare it with CompExp to show the importance of end-to-end policy gradient training.
    \item[-] \textbf{\model{}-BLEU}: our model trained with BLEU instead of IDF-BLEU. We create this variant to demonstrate the flaws of using BLEU to evaluate the quality of generated explanations.
\end{itemize}  

\begin{table*}[t]
    \caption{Explanation quality evaluated under IDF-BLEU, BLEU, average sentence length, average IDF per word, rep/l, seq\_rep\_2, feature precision and recall on RateBeer and TripAdvisor datasets. Bold numbers are the best of the corresponding metrics with \emph{p}-value < 0.05.}
    \vspace{-2mm}
    \setlength\tabcolsep{4.6pt}
    \label{tab:exp_eval}
    \begin{tabular}{|c|ccc|ccc|c|c|c|c|cc|}
        \hline
        \multirow{2}{*}{Model} & \multicolumn{3}{|c|}{IDF-BLEU} & \multicolumn{3}{|c|}{BLEU} & \multirow{2}{*}{Avg Length} & \multirow{2}{*}{IDF/word} & \multirow{2}{*}{rep/l} & \multirow{2}{*}{seq\_rep\_2} & \multicolumn{2}{|c|}{Feature} \\
        & 1 & 2 & 4 & 1 & 2 & 4 & & & & & precision & recall \\
        \hline
        \multicolumn{13}{|c|}{RateBeer} \\
        \hline

        Human & / & / & / & / & / & / & 11.13 & 2.45 & 0.0535 & 0.0015 & / & / \\
        NARRE & 17.00 & 5.18 & 1.29 & 30.22 & 9.90 & 3.58 & 11.50 & 2.43 & 0.0643 & 0.0013 & 0.2217 & 0.0722\\
        NRT & 30.38 & 16.30 & 5.80 & 48.22 & 25.28 & 10.03 & 10.43 & 2.09 & 0.1123 & 0.0240 & 0.4563 & 0.1320\\
        SAER & 31.79 & 16.02 & 5.71 & 49.08 & 26.87 & 10.59 & 10.71 & 1.93 & 0.1146 & 0.0223 & 0.4751 & 0.1347\\
        \model{}-Ext & 24.86 & 11.72 & 2.99 & 38.54 & 18.74 & 5.98 & 12.10 & 2.36 & 0.0420 & 0.0010 & 0.3092 & 0.0929\\
        \model{}-Pretrain & 27.59 & 13.44 & 4.19 & 44.93 & 21.53 & 7.95 & 10.55 & 2.07 & 0.1448 & 0.0381 & 0.3922 & 0.1123\\
        \model{}-BLEU & 23.20 & 14.55 & 4.70 & \textbf{53.45} & \textbf{32.42} & \textbf{11.62} & 7.09 & 1.83 & 0.0266 & 0.0006 & 0.4025 & 0.1173\\
        \model{} & \textbf{32.36} & \textbf{19.55} & \textbf{6.95} & 49.14 & 29.63 & 11.41 & 10.52 & 2.16 & 0.0572 & 0.0057 & \textbf{0.4796} & \textbf{0.1383}\\
        
        \hline
        \multicolumn{13}{|c|}{TripAdvisor} \\
        \hline
        
        Human & / & / & / & / & / & / & 12.85 & 2.45 & 0.0604 & 0.0021 & / & / \\
        NARRE & 11.97 & 3.43 & 1.59 & 20.45 & 6.23 & 3.38 & 13.17 & 2.41 & 0.0641 & 0.0022 & 0.1733 & 0.1258 \\
        NRT & 16.19 & 7.50 & \textbf{2.48} & 30.62 & 13.07 & 5.11 & 10.22 & 1.81 & 0.1277 & 0.0135 & 0.2939 & 0.1866 \\
        SAER & 16.37 & 7.65 & 2.35 & 31.20 & 13.51 & 4.94 & 10.08 & 1.71 & 0.1361 & 0.0141 & \textbf{0.3178} & \textbf{0.1961} \\
        \model{}-Ext & 13.52 & 4.25 & 1.14 & 22.12 & 7.30 & 2.66 & 14.70 & 2.39 & 0.0726 & 0.0037 & 0.2218 & 0.1553 \\
        \model{}-Pretrain & 14.50 & 6.11 & 1.99 & 27.14 & 11.12 & 4.32 & 10.79 & 1.92 & 0.1177 & 0.0250 & 0.2736 & 0.1597 \\
        \model{}-BLEU & 17.04 & 7.39 & 2.04 & \textbf{32.67} & \textbf{14.66} & \textbf{5.53} & 10.77 & 1.76 & 0.1597 & 0.0277 & 0.2332 & 0.1637 \\
        \model{} & \textbf{21.35} & \textbf{8.01} & 2.16 & 31.70 & 12.23 & 4.16 & 13.35 & 2.12 & 0.0654 & 0.0053 & 0.3155 & 0.1930 \\

        \hline
    \end{tabular}
    % \\\emph{*p}-value < 0.05
    \vspace{-1mm}
\end{table*}

\subsection{Quality of Generated Explanations}
To comprehensively study the quality of generated explanations, we employ different types of performance metrics, including IDF-BLEU-\{1, 2, 4\}, BLEU-\{1, 2, 4\}, average sentence length, average IDF per word, rep/l and seq\_rep\_2, and feature precision \& recall. Both rep/l and seq\_rep\_2 are proposed in \cite{welleck2019neural} to evaluate content repetition and higher values mean the content is more repetitive. Features are items' representative attributes that users usually care the most \cite{wang2018explainable, xian2021ex3, yang2021explanation}, e.g., ``pool'' in Figure \ref{fig:example}. The precision and recall measure if features mentioned in the generated explanations also appear in the user's ground-truth review.
%, which suggests models' ability to deliver personalized non-generic explanations.
We also include ground-truth review sentences as a reference baseline (labeled as ``Human'') to study the differences between human and algorithm generated content. 
The results are reported in Table \ref{tab:exp_eval}.

\subsubsection{IDF-BLEU over BLEU.}

While \model{}-BLEU topped every BLEU category on both datasets, \model{} also led almost all IDF-BLEU categories. This shows the effectiveness of our model design and the importance of directly optimizing the target evaluation metrics. 
To understand whether IDF-BLEU is a better metric than BLEU in evaluating the generated explanations, we should consider how the ``ground-truth'' content from real users look like, e.g., their average length and IDF/word, which suggest how much information is usually contained in a user-written sentence. As we can clearly notice that Avg Length and IDF/word in \model{}-BLEU are much smaller than Human. This suggests simply optimizing BLEU led to much shorter and less informative content. This follows our discussion before: BLEU encourage a model to generate less words and abuse common words to achieve high n-gram precision. \model{}-BLEU's low feature precision and recall also reflect its weakness in providing informative content. Therefore, the witnessed ``advantages'' of \model{}-BLEU in BLEU most likely come from shorter and more generic sentences, instead of really being closer to the ground-truth content. %On the other hand, \model{}-BLEU shows no advantages in IDF-BLEU. These results strongly suggest IDF-BLEU is a better metric than BLEU for evaluating content overlap in our scenario. 

\subsubsection{Advantages of \model{}.}

There is clear performance gap between the extraction-based solutions (NARRE, \model{}-Ext) and generation-based ones (NRT, SAER, \model{}). While generation-based solutions largely outperformed extraction-based ones in content overlapping with ground-truth (IDF-BLEU, BLEU, feature precision and recall), they were generally very different from human writings in terms of sentence length, use of rare words (IDF/word), and content repetition (rep/l, seq\_res\_2). 
%This distinction well demonstrates the strengths and weaknesses of each category of explanation generation methods. 
The extraction-based solutions use content provided by human, but they are limited to the existing content. The generation-based solutions customize content for each recommendation, but suffer from common flaws of generative models, e.g., short, dull, and repetitive. Among all the models, \model{} achieved the best balance among all metrics. It significantly exceeded all baselines in terms of IDF-BLEU-\{1,2\} and its BLEU was only behind CompExp-BLEU. Its feature precision and recall are competitive with SAER while leading the rest, though SAER enjoys additional advantage from predefined feature pool of each item as input. As a generation-based model, \model{} largely improved the average length, word rarity, and reduced repetition over NRT and SAER. The only exception is that \model{}-BLEU was less repetitive in RateBeer, but it is mainly because its explanations were very short in general. 

\subsubsection{Ablation Study.}

Though \model{} performed well as a whole, it is inspiring to study if each component works as expected in the extract-and-refine workflow. 
%To check the effectiveness of modeling comparative explanations in the extractor, we compare IDF-BLEU and BLEU between .
First, both \model{}-Ext and NARRE are extraction-based with the same candidate pool, but \model{}-Ext showed obvious advantage under most categories of IDF-BLEU and BLEU. 
%This indicates that other comparative explanations of the user are indeed informative to the estimation of the target explanation. 
It suggests our extractor alone can act as a competent solution where generation-based models do not fit, e.g., real-time applications requiring minimum response time. 
% Interestingly, the extractor prefers sentences longer than NARRE and human's average. This imp
The comparison between \model{}-Ext and \model{}-Pretrain demonstrates that the refiner is able to leverage the gradient direction to improve the prototypes, even when the prototypes are given by an extractor that has not been trained jointly with the refiner. At last, there are huge gaps in all metrics between \model{}-Pretrain and \model{} in both datasets. It is obvious that our reward design is beneficial to both quality and diversity of the generated explanations.

\subsubsection{Comparativeness.}
% \noindent\textbf{Comparativeness.}
To verify if the generated explanations by \model{} capture the comparative ranking of items, we study its its output's sensitivity to the input recommendation ratings. As a starting point, the ground-truth explanation perfectly aligns with the recommendation ranking, which is derived from the ground-truth rating. If the generated explanation carries the same ranking of item, the generated content should be close to the ground-truth content. 
As a result, if we manipulate the input recommendation scores of items, the generated explanations should start to deviate. The further we push the rankings apart, the further the generated explanation should be pushed away from the ground-truth explanation. We use IDF-BLEU and BLEU to measure the content similarity and perturb the recommendation ratings with Gaussian noise. 
%The decreasing ratios of the similarity metrics over the standard deviation of the Gaussian noise are plotted in . 
As shown in Figure \ref{fig:noise}, all IDF-BLEU and BLEU metrics keep decreasing with the increasing amount of perturbation. In other words, even if it is for the same user and same set of items, with different recommendation scores assigned, \model{} would generate different explanations to explain their relative ranking. %Moreover, the decreasing ratio of every IDF-BLEU category is constantly higher than its counterpart of BLEU. This observation shows the advantage of IDF-BLEU over BLEU that IDF-BLEU is a more sensitive metric to reflect the comparative ranking in explanations. 

\subsubsection{Predicted Ratings.}
\begin{figure}[t]
    \centering
    \begin{subfigure}[b]{0.5\linewidth}
      \centering
      \includegraphics[width=\linewidth]{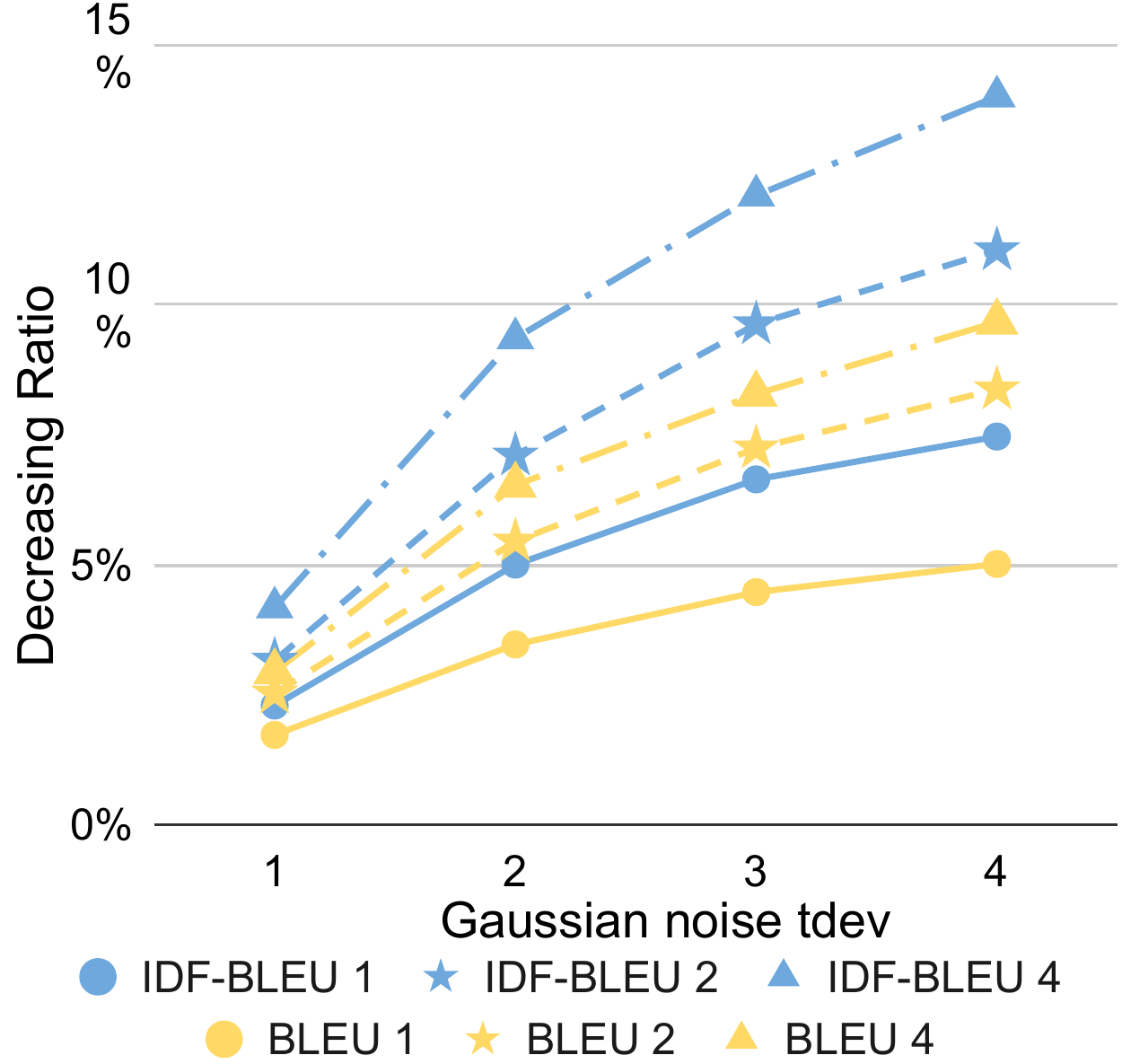}
      \caption{}
      \label{fig:noise}
        \vspace{-4mm}
    \end{subfigure}%
    \begin{subfigure}[b]{.5\linewidth}
      \centering
      \includegraphics[width=\linewidth]{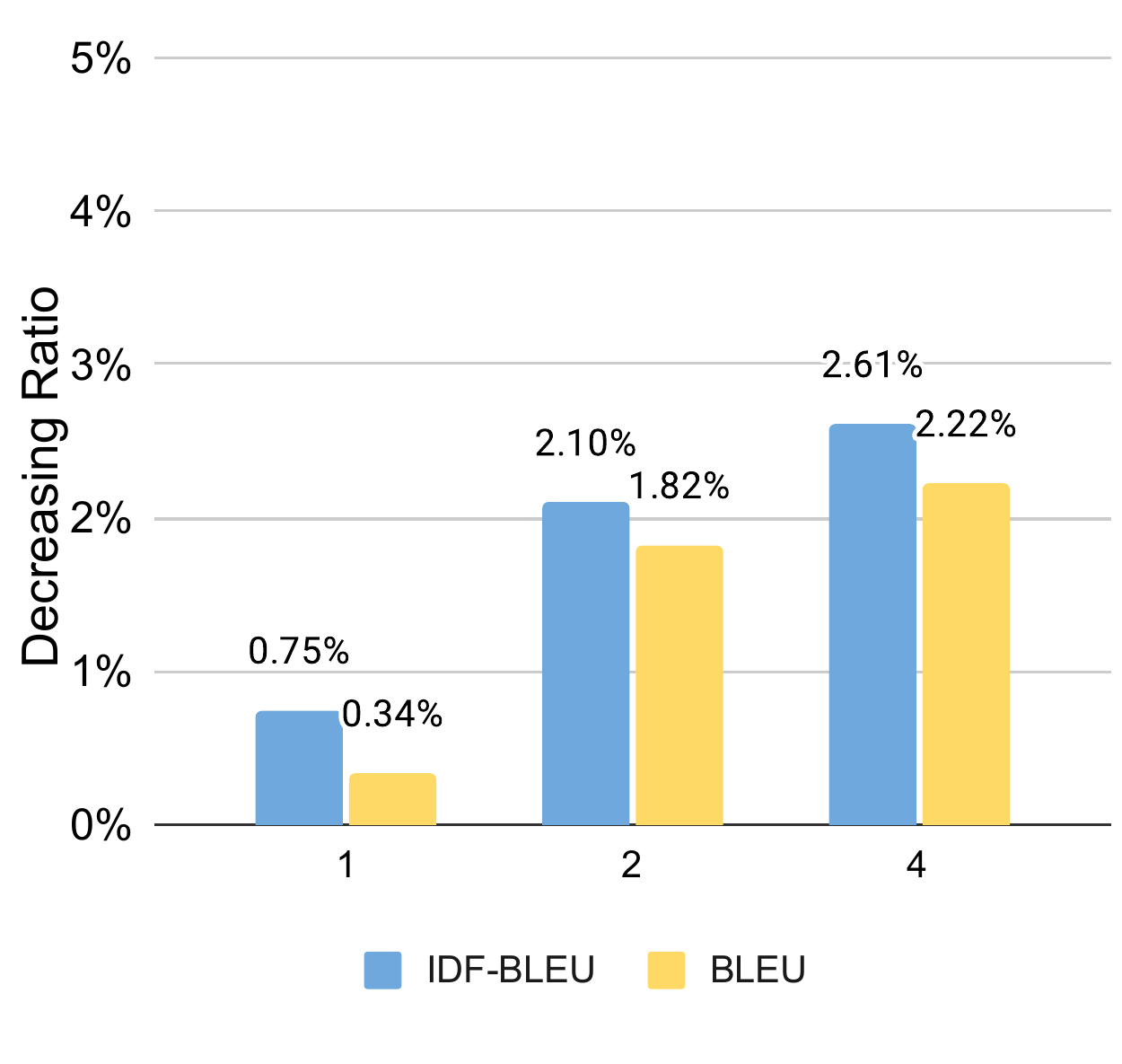}
      \caption{}
      \label{fig:predicted}
          \vspace{-4mm}
    \end{subfigure}
    \caption{(a) Impact of noise in recommendation ratings on BLEU and IDF-BLEU. (b) Change in BLEU and IDF-BLEU with algorithm's predicted ratings.}
    \label{fig:test}
    \vspace{-4mm}
\end{figure}

% \noindent\textbf{Predicted Ratings.} 
Motivated by the findings in Figure \ref{fig:noise}, we further study how \model{} is influenced by a real recommendation algorithm's predicted ratings. We employed the neural collaborative filtering \cite{he2017neural} and used its predicted ratings in \model{}'s training and testing. The result is plotted in Figure \ref{fig:predicted}. Compared with previous randomly perturbed ratings, the predicted ratings bring very limited changes to the explanations. This confirms our experiment results based on ground-truth ratings can fairly represent \model{}'s performance in real-world usage scenarios.

% \begin{figure}[t]
%     \centering
%     \includegraphics[width=0.8\linewidth]{figures/noise.png}
%     \caption{BLEU and IDF-BLEU's Decreasing ratio over Gaussian noise on recommendation rating.}
%     \Description{BLEU and IDF-BLEU's Decreasing ratio over Gaussian noise on recommendation rating.}
%     \label{fig:noise}
%     \vspace{-4mm}
% \end{figure}

\section{User Study}

We have three research questions to answer in user study: 1) does users' judgement toward explanation quality aligns more with IDF-BLEU than BLEU; 2) do users find our comparative explanations more helpful than the baselines'; and 3) can users better perceive the comparative ranking from our explanations than the baselines'. To answer these three research questions, we design two user study tasks based on RateBeer dataset using Amazon Mechanical Turk.

The first task studies the first two research questions together.
Specifically, we shuffle explanations from different models about the same recommended item and ask the participants to compare them, and then select the most helpful ones. 
To help participants evaluate the explanation quality, we include the original user review as the item description, towards which they can judge if the explanation are accurate or informative.  
For each recommended item, we ask participants to answer the following question after reading its description and candidate explanations:
\begin{itemize}[leftmargin=*]
\item[]  \textit{``Which of the following explanations best describe the characteristics of the given beer and help you the most to understand why you should pay attention to the recommendation?''}
\end{itemize}
% Participants can choose multiple explanations if they feel they are equally good. 
In this experiment, 
% we compared explanations from \model{}, SAER, NRT, and NARRE; and in the end, 
we collected 660 user responses.

The results are presented in Table \ref{tab:kappa} and \ref{tab:upvote}. In Table \ref{tab:kappa}, we used Cohen's kappa coefficient to compare IDF-BLEU and BLEU's agreement with users' responses. 
% Since it is about comparison between IDF-BLEU and BLEU, the source model does not matter. 
For each test case, we pair explanations that the participants chose as helpful with the rest to form a set of explanation pairs. Then we use IDF-BLEU-\{1,2,4\} and BLEU-\{1,2,4\} to identify the helpful one in each pair. The kappa coefficient shows that IDF-BLEU aligns significantly better with users' judgment in all three subcategories under paired t-test. %This result once again demonstrates IDF-BLEU is a more appropriate metric for explanation evaluation. 
Table \ref{tab:upvote} shows the helpfulness vote on each model and the paired t-test results of \model{} against other baselines. The helpfulness vote on \model{} is significantly higher than others, which suggests strong user preference over its generated explanations.  

\begin{table}[t]
    \caption{Cohen's kappa coefficient of explanation quality between the human judgements and BLEU \& IDF-BLEU.}
    \vspace{-2mm}
    \label{tab:kappa}
    \begin{tabular}{|c|c|ccc|}
    \hline
    \multicolumn{2}{|c|}{} & 1 & 2 & 4 \\
    \cline{1-5}
    \multirow{2}{*}{$\kappa$} & BLEU & 0.2936 & 0.3114 & 0.2814 \\
     & IDF-BLEU & 0.3452 & 0.3396 & 0.3152 \\
    \hline
    \multicolumn{2}{|c|}{Paired t-test} & 0.0001 & 0.0094 & 0.0071 \\
    \hline
\end{tabular}
\vspace{-2mm}
\end{table}

\begin{table}[t]
    \caption{Up-vote rate of explanations' helpfulness.}
    \vspace{-2mm}
    \label{tab:upvote}
    \begin{tabular}{|c|cccc|}
    \hline
      & \model{} & SAER & NRT & NARRE \\
    \hline
    Up-vote Rate & 43.79\% & 37.27\% & 35.61\% & 30.61\% \\
    \hline
    % Paired t-test & \multirow{2}{*}{/} & \multirow{2}{*}{0.0182} & \multirow{2}{*}{0.009} & \multirow{2}{*}{0} \\
    % (\model{} vs .) & & & & \\
    Paired t-test & / & 0.0182 & 0.009 & 0 \\
    \hline
    \end{tabular}
    \vspace{-2mm}
\end{table}

The second task addresses the last research question, i.e., if a user is able to perceive the ranking of recommended items from the explanations. In this task, we randomly paired items of different ratings and asked participants to identify which item is better by reading the provided explanations. We then evaluated the agreement rate between participants' choices and the actual ranking. 
In particular, given the explanations of a model, the participants were required to answer the following question:
\begin{itemize}[leftmargin=*]
\item[]  \textit{``After reading the explanations for recommended items, which item would you like to choose? You are expected to judge the quality of the items based on the provided explanations.''}
\end{itemize}
We chose SAER and NRT as baselines. Besides, we also include the ground-truth sentences from the actual user reviews as a reference. We collected 200 responses for each model. 

\begin{table}[t]
    \caption{Agreement rate between actual ranking and the users perceived ranking of paired items based on the provided explanations.}
    \vspace{-2mm}
    \label{tab:ar}
    \begin{tabular}{|c|cccc|}
    \hline
     & GT & \model{} & SAER & NRT \\
    \hline
    Agreement Rate & 72.29\% & 57.27\% & 56.25\% & 53.14\% \\
    \hline
    \end{tabular}
    \vspace{-2mm}
\end{table}

Table \ref{tab:ar} reports the agreement rates between the actual ranking and the ranking perceived by the participants. \model{}'s agreement rate is slightly higher than NRT and SAER, but it is far below the Ground-Truth. The Ground-Truth's high agreement rate quantitatively confirms that the original user provided review sentences are highly comparative. This observation supports our choice of training the comparative explanation generation from paired user review sentences. And it also suggests there is still a performance gap in comparativeness for learning-based solutions to bridge. And an improved objective for optimization, e.g., include quantified pairwise comparativeness, might be a promising direction.
%reveals that although \model{} does better help users perceive the recommendation rankings, it has not yet met human's expectation of comparative explanations.
\section{Conclusion and Future Work}
In this paper, we studied the problem of comparative explanation generation in explainable recommendation. The objective of our generated explanations is to help users understand the comparative item rankings provided in a recommender system. 
%Ideally, after reading the provided explanations, the user should reach the same conclusion of item ranking as provided by the system. 
We develop a neural extract-and-refine architecture to generate such comparative explanations, with customized metrics to penalize generic and useless content in the generated explanations. Both offline evaluations and user studies demonstrated the effectiveness of our solution.

This work starts a bright new direction in explainable recommendation. Our current solution only focuses on explanation generation, by assuming a perfect recommendation algorithm (i.e., we directly used the ground-truth opinion ratings in our experiments). It is important to improve our model by co-design with a real recommendation algorithm, whose recommendation score is expected to be noise and erroneous. In addition, we still heavily depend on existing review content to guide explanation generation. It will be more meaningful to introduce actual user feedback in this process, i.e., interactive optimization of explanation generation.

\begin{acks}
We thank the anonymous reviewers for their insightful comments and suggestions. This work is partially supported by the National Science Foundation under grant SCH-1838615, IIS-1553568, and IIS-2007492, and by Alibaba Group through Alibaba Innovative Research Program. 
\end{acks}

%%
%% The next two lines define the bibliography style to be used, and
%% the bibliography file.
\bibliographystyle{ACM-Reference-Format}
\bibliography{main}

%%% -*-BibTeX-*-
%%% Do NOT edit. File created by BibTeX with style
%%% ACM-Reference-Format-Journals [18-Jan-2012].

\begin{thebibliography}{48}

%%% ====================================================================
%%% NOTE TO THE USER: you can override these defaults by providing
%%% customized versions of any of these macros before the \bibliography
%%% command.  Each of them MUST provide its own final punctuation,
%%% except for \shownote{}, \showDOI{}, and \showURL{}.  The latter two
%%% do not use final punctuation, in order to avoid confusing it with
%%% the Web address.
%%%
%%% To suppress output of a particular field, define its macro to expand
%%% to an empty string, or better, \unskip, like this:
%%%
%%% \newcommand{\showDOI}[1]{\unskip}   % LaTeX syntax
%%%
%%% \def \showDOI #1{\unskip}           % plain TeX syntax
%%%
%%% ====================================================================

\ifx \showCODEN    \undefined \def \showCODEN     #1{\unskip}     \fi
\ifx \showDOI      \undefined \def \showDOI       #1{#1}\fi
\ifx \showISBNx    \undefined \def \showISBNx     #1{\unskip}     \fi
\ifx \showISBNxiii \undefined \def \showISBNxiii  #1{\unskip}     \fi
\ifx \showISSN     \undefined \def \showISSN      #1{\unskip}     \fi
\ifx \showLCCN     \undefined \def \showLCCN      #1{\unskip}     \fi
\ifx \shownote     \undefined \def \shownote      #1{#1}          \fi
\ifx \showarticletitle \undefined \def \showarticletitle #1{#1}   \fi
\ifx \showURL      \undefined \def \showURL       {\relax}        \fi
% The following commands are used for tagged output and should be
% invisible to TeX
\providecommand\bibfield[2]{#2}
\providecommand\bibinfo[2]{#2}
\providecommand\natexlab[1]{#1}
\providecommand\showeprint[2][]{arXiv:#2}

\bibitem[\protect\citeauthoryear{Aggarwal et~al\mbox{.}}{Aggarwal
  et~al\mbox{.}}{2016}]%
        {aggarwal2016recommender}
\bibfield{author}{\bibinfo{person}{Charu~C Aggarwal} {et~al\mbox{.}}}
  \bibinfo{year}{2016}\natexlab{}.
\newblock \bibinfo{booktitle}{\emph{Recommender systems}}.
  Vol.~\bibinfo{volume}{1}.
\newblock \bibinfo{publisher}{Springer}.
\newblock


\bibitem[\protect\citeauthoryear{Ai, Azizi, Chen, and Zhang}{Ai
  et~al\mbox{.}}{2018}]%
        {ai2018learning}
\bibfield{author}{\bibinfo{person}{Qingyao Ai}, \bibinfo{person}{Vahid Azizi},
  \bibinfo{person}{Xu Chen}, {and} \bibinfo{person}{Yongfeng Zhang}.}
  \bibinfo{year}{2018}\natexlab{}.
\newblock \showarticletitle{Learning heterogeneous knowledge base embeddings
  for explainable recommendation}.
\newblock \bibinfo{journal}{\emph{Algorithms}} \bibinfo{volume}{11},
  \bibinfo{number}{9} (\bibinfo{year}{2018}), \bibinfo{pages}{137}.
\newblock


\bibitem[\protect\citeauthoryear{Balabanovi{\'c} and Shoham}{Balabanovi{\'c}
  and Shoham}{1997}]%
        {balabanovic1997fab}
\bibfield{author}{\bibinfo{person}{Marko Balabanovi{\'c}} {and}
  \bibinfo{person}{Yoav Shoham}.} \bibinfo{year}{1997}\natexlab{}.
\newblock \showarticletitle{Fab: content-based, collaborative recommendation}.
\newblock \bibinfo{journal}{\emph{Commun. ACM}} \bibinfo{volume}{40},
  \bibinfo{number}{3} (\bibinfo{year}{1997}), \bibinfo{pages}{66--72}.
\newblock


\bibitem[\protect\citeauthoryear{Bilgic and Mooney}{Bilgic and Mooney}{2005}]%
        {bilgic2005explaining}
\bibfield{author}{\bibinfo{person}{Mustafa Bilgic} {and}
  \bibinfo{person}{Raymond~J Mooney}.} \bibinfo{year}{2005}\natexlab{}.
\newblock \showarticletitle{Explaining recommendations: Satisfaction vs.
  promotion}. In \bibinfo{booktitle}{\emph{Beyond Personalization Workshop,
  IUI}}, Vol.~\bibinfo{volume}{5}.
\newblock


\bibitem[\protect\citeauthoryear{Cai, Wang, and Wang}{Cai
  et~al\mbox{.}}{2017}]%
        {cai2017accounting}
\bibfield{author}{\bibinfo{person}{Renqin Cai}, \bibinfo{person}{Chi Wang},
  {and} \bibinfo{person}{Hongning Wang}.} \bibinfo{year}{2017}\natexlab{}.
\newblock \showarticletitle{Accounting for the Correspondence in Commented
  Data}. In \bibinfo{booktitle}{\emph{Proceedings of the 40th International ACM
  SIGIR Conference on Research and Development in Information Retrieval}}.
  \bibinfo{pages}{365--374}.
\newblock


\bibitem[\protect\citeauthoryear{Cai, Wu, San, Wang, and Wang}{Cai
  et~al\mbox{.}}{2021}]%
        {cai2021category}
\bibfield{author}{\bibinfo{person}{Renqin Cai}, \bibinfo{person}{Jibang Wu},
  \bibinfo{person}{Aidan San}, \bibinfo{person}{Chong Wang}, {and}
  \bibinfo{person}{Hongning Wang}.} \bibinfo{year}{2021}\natexlab{}.
\newblock \showarticletitle{Category-aware collaborative sequential
  recommendation}. In \bibinfo{booktitle}{\emph{Proceedings of the 44th
  International ACM SIGIR Conference on Research and Development in Information
  Retrieval}}. \bibinfo{pages}{388--397}.
\newblock


\bibitem[\protect\citeauthoryear{Chen, Zhang, Liu, and Ma}{Chen
  et~al\mbox{.}}{2018}]%
        {chen2018neural}
\bibfield{author}{\bibinfo{person}{Chong Chen}, \bibinfo{person}{Min Zhang},
  \bibinfo{person}{Yiqun Liu}, {and} \bibinfo{person}{Shaoping Ma}.}
  \bibinfo{year}{2018}\natexlab{}.
\newblock \showarticletitle{Neural attentional rating regression with
  review-level explanations}. In \bibinfo{booktitle}{\emph{Proceedings of the
  2018 World Wide Web Conference}}. \bibinfo{pages}{1583--1592}.
\newblock


\bibitem[\protect\citeauthoryear{Chen, Li, Sun, Xu, and Yin}{Chen
  et~al\mbox{.}}{2021}]%
        {chen2021temporal}
\bibfield{author}{\bibinfo{person}{Hongxu Chen}, \bibinfo{person}{Yicong Li},
  \bibinfo{person}{Xiangguo Sun}, \bibinfo{person}{Guandong Xu}, {and}
  \bibinfo{person}{Hongzhi Yin}.} \bibinfo{year}{2021}\natexlab{}.
\newblock \showarticletitle{Temporal meta-path guided explainable
  recommendation}. In \bibinfo{booktitle}{\emph{Proceedings of the 14th ACM
  International Conference on Web Search and Data Mining}}.
  \bibinfo{pages}{1056--1064}.
\newblock


\bibitem[\protect\citeauthoryear{Chen, Yin, Ye, Huang, Wang, and Wang}{Chen
  et~al\mbox{.}}{2020}]%
        {chen2020try}
\bibfield{author}{\bibinfo{person}{Tong Chen}, \bibinfo{person}{Hongzhi Yin},
  \bibinfo{person}{Guanhua Ye}, \bibinfo{person}{Zi Huang},
  \bibinfo{person}{Yang Wang}, {and} \bibinfo{person}{Meng Wang}.}
  \bibinfo{year}{2020}\natexlab{}.
\newblock \showarticletitle{Try this instead: Personalized and interpretable
  substitute recommendation}. In \bibinfo{booktitle}{\emph{Proceedings of the
  43rd International ACM SIGIR Conference on Research and Development in
  Information Retrieval}}. \bibinfo{pages}{891--900}.
\newblock


\bibitem[\protect\citeauthoryear{Chung, Gulcehre, Cho, and Bengio}{Chung
  et~al\mbox{.}}{2014}]%
        {chung2014empirical}
\bibfield{author}{\bibinfo{person}{Junyoung Chung}, \bibinfo{person}{Caglar
  Gulcehre}, \bibinfo{person}{KyungHyun Cho}, {and} \bibinfo{person}{Yoshua
  Bengio}.} \bibinfo{year}{2014}\natexlab{}.
\newblock \showarticletitle{Empirical evaluation of gated recurrent neural
  networks on sequence modeling}.
\newblock \bibinfo{journal}{\emph{arXiv preprint arXiv:1412.3555}}
  (\bibinfo{year}{2014}).
\newblock


\bibitem[\protect\citeauthoryear{Conneau, Kiela, Schwenk, Barrault, and
  Bordes}{Conneau et~al\mbox{.}}{2017}]%
        {conneau2017supervised}
\bibfield{author}{\bibinfo{person}{Alexis Conneau}, \bibinfo{person}{Douwe
  Kiela}, \bibinfo{person}{Holger Schwenk}, \bibinfo{person}{Lo{\"\i}c
  Barrault}, {and} \bibinfo{person}{Antoine Bordes}.}
  \bibinfo{year}{2017}\natexlab{}.
\newblock \showarticletitle{Supervised Learning of Universal Sentence
  Representations from Natural Language Inference Data}. In
  \bibinfo{booktitle}{\emph{Proceedings of the 2017 Conference on Empirical
  Methods in Natural Language Processing}}. \bibinfo{pages}{670--680}.
\newblock


\bibitem[\protect\citeauthoryear{Guu, Hashimoto, Oren, and Liang}{Guu
  et~al\mbox{.}}{2018}]%
        {guu2018generating}
\bibfield{author}{\bibinfo{person}{Kelvin Guu}, \bibinfo{person}{Tatsunori~B
  Hashimoto}, \bibinfo{person}{Yonatan Oren}, {and} \bibinfo{person}{Percy
  Liang}.} \bibinfo{year}{2018}\natexlab{}.
\newblock \showarticletitle{Generating sentences by editing prototypes}.
\newblock \bibinfo{journal}{\emph{Transactions of the Association for
  Computational Linguistics}}  \bibinfo{volume}{6} (\bibinfo{year}{2018}),
  \bibinfo{pages}{437--450}.
\newblock


\bibitem[\protect\citeauthoryear{He, Chen, Kan, and Chen}{He
  et~al\mbox{.}}{2015}]%
        {he2015trirank}
\bibfield{author}{\bibinfo{person}{Xiangnan He}, \bibinfo{person}{Tao Chen},
  \bibinfo{person}{Min-Yen Kan}, {and} \bibinfo{person}{Xiao Chen}.}
  \bibinfo{year}{2015}\natexlab{}.
\newblock \showarticletitle{Trirank: Review-aware explainable recommendation by
  modeling aspects}. In \bibinfo{booktitle}{\emph{Proceedings of the 24th ACM
  International on Conference on Information and Knowledge Management}}.
  \bibinfo{pages}{1661--1670}.
\newblock


\bibitem[\protect\citeauthoryear{He, Liao, Zhang, Nie, Hu, and Chua}{He
  et~al\mbox{.}}{2017}]%
        {he2017neural}
\bibfield{author}{\bibinfo{person}{Xiangnan He}, \bibinfo{person}{Lizi Liao},
  \bibinfo{person}{Hanwang Zhang}, \bibinfo{person}{Liqiang Nie},
  \bibinfo{person}{Xia Hu}, {and} \bibinfo{person}{Tat-Seng Chua}.}
  \bibinfo{year}{2017}\natexlab{}.
\newblock \showarticletitle{Neural collaborative filtering}. In
  \bibinfo{booktitle}{\emph{Proceedings of the 26th international conference on
  world wide web}}. \bibinfo{pages}{173--182}.
\newblock


\bibitem[\protect\citeauthoryear{Herlocker, Konstan, and Riedl}{Herlocker
  et~al\mbox{.}}{2000}]%
        {herlocker2000explaining}
\bibfield{author}{\bibinfo{person}{Jonathan~L Herlocker},
  \bibinfo{person}{Joseph~A Konstan}, {and} \bibinfo{person}{John Riedl}.}
  \bibinfo{year}{2000}\natexlab{}.
\newblock \showarticletitle{Explaining collaborative filtering
  recommendations}. In \bibinfo{booktitle}{\emph{Proceedings of the 2000 ACM
  conference on Computer supported cooperative work}}. ACM,
  \bibinfo{pages}{241--250}.
\newblock


\bibitem[\protect\citeauthoryear{Holtzman, Buys, Du, Forbes, and Choi}{Holtzman
  et~al\mbox{.}}{2019}]%
        {holtzman2019curious}
\bibfield{author}{\bibinfo{person}{Ari Holtzman}, \bibinfo{person}{Jan Buys},
  \bibinfo{person}{Li Du}, \bibinfo{person}{Maxwell Forbes}, {and}
  \bibinfo{person}{Yejin Choi}.} \bibinfo{year}{2019}\natexlab{}.
\newblock \showarticletitle{The curious case of neural text degeneration}.
\newblock \bibinfo{journal}{\emph{arXiv preprint arXiv:1904.09751}}
  (\bibinfo{year}{2019}).
\newblock


\bibitem[\protect\citeauthoryear{Ji and Shen}{Ji and Shen}{2016}]%
        {ji2016jointly}
\bibfield{author}{\bibinfo{person}{Ke Ji} {and} \bibinfo{person}{Hong Shen}.}
  \bibinfo{year}{2016}\natexlab{}.
\newblock \showarticletitle{Jointly modeling content, social network and
  ratings for explainable and cold-start recommendation}.
\newblock \bibinfo{journal}{\emph{Neurocomputing}}  \bibinfo{volume}{218}
  (\bibinfo{year}{2016}), \bibinfo{pages}{1--12}.
\newblock


\bibitem[\protect\citeauthoryear{Karatzoglou, Baltrunas, and Shi}{Karatzoglou
  et~al\mbox{.}}{2013}]%
        {karatzoglou2013learning}
\bibfield{author}{\bibinfo{person}{Alexandros Karatzoglou},
  \bibinfo{person}{Linas Baltrunas}, {and} \bibinfo{person}{Yue Shi}.}
  \bibinfo{year}{2013}\natexlab{}.
\newblock \showarticletitle{Learning to rank for recommender systems}. In
  \bibinfo{booktitle}{\emph{Proceedings of the 7th ACM Conference on
  Recommender Systems}}. \bibinfo{pages}{493--494}.
\newblock


\bibitem[\protect\citeauthoryear{Koren, Bell, and Volinsky}{Koren
  et~al\mbox{.}}{2009}]%
        {koren2009matrix}
\bibfield{author}{\bibinfo{person}{Yehuda Koren}, \bibinfo{person}{Robert
  Bell}, {and} \bibinfo{person}{Chris Volinsky}.}
  \bibinfo{year}{2009}\natexlab{}.
\newblock \showarticletitle{Matrix factorization techniques for recommender
  systems}.
\newblock \bibinfo{journal}{\emph{Computer}} \bibinfo{volume}{42},
  \bibinfo{number}{8} (\bibinfo{year}{2009}), \bibinfo{pages}{30--37}.
\newblock


\bibitem[\protect\citeauthoryear{Li, Quan, Peng, Qi, Deng, and Wu}{Li
  et~al\mbox{.}}{2019}]%
        {li2019capsule}
\bibfield{author}{\bibinfo{person}{Chenliang Li}, \bibinfo{person}{Cong Quan},
  \bibinfo{person}{Li Peng}, \bibinfo{person}{Yunwei Qi},
  \bibinfo{person}{Yuming Deng}, {and} \bibinfo{person}{Libing Wu}.}
  \bibinfo{year}{2019}\natexlab{}.
\newblock \showarticletitle{A Capsule Network for Recommendation and Explaining
  What You Like and Dislike}. In \bibinfo{booktitle}{\emph{Proceedings of the
  42nd International ACM SIGIR Conference on Research and Development in
  Information Retrieval}}. \bibinfo{pages}{275--284}.
\newblock


\bibitem[\protect\citeauthoryear{Li, Wang, Ren, Bing, and Lam}{Li
  et~al\mbox{.}}{2017}]%
        {li2017neural}
\bibfield{author}{\bibinfo{person}{Piji Li}, \bibinfo{person}{Zihao Wang},
  \bibinfo{person}{Zhaochun Ren}, \bibinfo{person}{Lidong Bing}, {and}
  \bibinfo{person}{Wai Lam}.} \bibinfo{year}{2017}\natexlab{}.
\newblock \showarticletitle{Neural rating regression with abstractive tips
  generation for recommendation}. In \bibinfo{booktitle}{\emph{Proceedings of
  the 40th International ACM SIGIR conference on Research and Development in
  Information Retrieval}}. \bibinfo{pages}{345--354}.
\newblock


\bibitem[\protect\citeauthoryear{Lundberg and Lee}{Lundberg and Lee}{2017}]%
        {lundberg2017unified}
\bibfield{author}{\bibinfo{person}{Scott~M Lundberg} {and}
  \bibinfo{person}{Su-In Lee}.} \bibinfo{year}{2017}\natexlab{}.
\newblock \showarticletitle{A unified approach to interpreting model
  predictions}. In \bibinfo{booktitle}{\emph{Advances in neural information
  processing systems}}. \bibinfo{pages}{4765--4774}.
\newblock


\bibitem[\protect\citeauthoryear{Luong, Pham, and Manning}{Luong
  et~al\mbox{.}}{2015}]%
        {luong2015effective}
\bibfield{author}{\bibinfo{person}{Minh-Thang Luong}, \bibinfo{person}{Hieu
  Pham}, {and} \bibinfo{person}{Christopher~D Manning}.}
  \bibinfo{year}{2015}\natexlab{}.
\newblock \showarticletitle{Effective approaches to attention-based neural
  machine translation}.
\newblock \bibinfo{journal}{\emph{arXiv preprint arXiv:1508.04025}}
  (\bibinfo{year}{2015}).
\newblock


\bibitem[\protect\citeauthoryear{McAuley, Leskovec, and Jurafsky}{McAuley
  et~al\mbox{.}}{2012}]%
        {julian2012learning}
\bibfield{author}{\bibinfo{person}{Julian McAuley}, \bibinfo{person}{Jure
  Leskovec}, {and} \bibinfo{person}{Dan Jurafsky}.}
  \bibinfo{year}{2012}\natexlab{}.
\newblock \showarticletitle{Learning Attitudes and Attributes from Multi-Aspect
  Reviews}. In \bibinfo{booktitle}{\emph{Proceedings of the 2012 IEEE 12th
  International Conference on Data Mining}} \emph{(\bibinfo{series}{ICDM
  ’12})}. \bibinfo{publisher}{IEEE Computer Society}, \bibinfo{address}{USA},
  \bibinfo{pages}{1020–1025}.
\newblock
\showISBNx{9780769549057}


\bibitem[\protect\citeauthoryear{McAuley, Pandey, and Leskovec}{McAuley
  et~al\mbox{.}}{2015}]%
        {mcauley2015inferring}
\bibfield{author}{\bibinfo{person}{Julian McAuley}, \bibinfo{person}{Rahul
  Pandey}, {and} \bibinfo{person}{Jure Leskovec}.}
  \bibinfo{year}{2015}\natexlab{}.
\newblock \showarticletitle{Inferring networks of substitutable and
  complementary products}. In \bibinfo{booktitle}{\emph{Proceedings of the 21th
  ACM SIGKDD international conference on knowledge discovery and data mining}}.
  \bibinfo{pages}{785--794}.
\newblock


\bibitem[\protect\citeauthoryear{Papineni, Roukos, Ward, and Zhu}{Papineni
  et~al\mbox{.}}{2002}]%
        {papineni2002bleu}
\bibfield{author}{\bibinfo{person}{Kishore Papineni}, \bibinfo{person}{Salim
  Roukos}, \bibinfo{person}{Todd Ward}, {and} \bibinfo{person}{Wei-Jing Zhu}.}
  \bibinfo{year}{2002}\natexlab{}.
\newblock \showarticletitle{BLEU: a method for automatic evaluation of machine
  translation}. In \bibinfo{booktitle}{\emph{Proceedings of the 40th annual
  meeting on association for computational linguistics}}. Association for
  Computational Linguistics, \bibinfo{pages}{311--318}.
\newblock


\bibitem[\protect\citeauthoryear{Pennington, Socher, and Manning}{Pennington
  et~al\mbox{.}}{2014}]%
        {pennington2014glove}
\bibfield{author}{\bibinfo{person}{Jeffrey Pennington},
  \bibinfo{person}{Richard Socher}, {and} \bibinfo{person}{Christopher~D
  Manning}.} \bibinfo{year}{2014}\natexlab{}.
\newblock \showarticletitle{Glove: Global vectors for word representation}. In
  \bibinfo{booktitle}{\emph{Proceedings of the 2014 conference on empirical
  methods in natural language processing (EMNLP)}}.
  \bibinfo{pages}{1532--1543}.
\newblock


\bibitem[\protect\citeauthoryear{Rendle}{Rendle}{2010}]%
        {rendle2010factorization}
\bibfield{author}{\bibinfo{person}{Steffen Rendle}.}
  \bibinfo{year}{2010}\natexlab{}.
\newblock \showarticletitle{Factorization machines}. In
  \bibinfo{booktitle}{\emph{2010 IEEE International Conference on Data
  Mining}}. IEEE, \bibinfo{pages}{995--1000}.
\newblock


\bibitem[\protect\citeauthoryear{Rendle, Freudenthaler, Gantner, and
  Schmidt-Thieme}{Rendle et~al\mbox{.}}{2012}]%
        {rendle2012bpr}
\bibfield{author}{\bibinfo{person}{Steffen Rendle}, \bibinfo{person}{Christoph
  Freudenthaler}, \bibinfo{person}{Zeno Gantner}, {and} \bibinfo{person}{Lars
  Schmidt-Thieme}.} \bibinfo{year}{2012}\natexlab{}.
\newblock \showarticletitle{BPR: Bayesian personalized ranking from implicit
  feedback}.
\newblock \bibinfo{journal}{\emph{arXiv preprint arXiv:1205.2618}}
  (\bibinfo{year}{2012}).
\newblock


\bibitem[\protect\citeauthoryear{Ribeiro, Singh, and Guestrin}{Ribeiro
  et~al\mbox{.}}{2016}]%
        {ribeiro2016should}
\bibfield{author}{\bibinfo{person}{Marco~Tulio Ribeiro},
  \bibinfo{person}{Sameer Singh}, {and} \bibinfo{person}{Carlos Guestrin}.}
  \bibinfo{year}{2016}\natexlab{}.
\newblock \showarticletitle{"Why should i trust you?" Explaining the
  predictions of any classifier}. In \bibinfo{booktitle}{\emph{Proceedings of
  the 22nd ACM SIGKDD international conference on knowledge discovery and data
  mining}}. \bibinfo{pages}{1135--1144}.
\newblock


\bibitem[\protect\citeauthoryear{Sarwar, Karypis, Konstan, and Riedl}{Sarwar
  et~al\mbox{.}}{2001}]%
        {sarwar2001item}
\bibfield{author}{\bibinfo{person}{Badrul Sarwar}, \bibinfo{person}{George
  Karypis}, \bibinfo{person}{Joseph Konstan}, {and} \bibinfo{person}{John
  Riedl}.} \bibinfo{year}{2001}\natexlab{}.
\newblock \showarticletitle{Item-based collaborative filtering recommendation
  algorithms}. In \bibinfo{booktitle}{\emph{Proceedings of the 10th
  international conference on World Wide Web}}. \bibinfo{pages}{285--295}.
\newblock


\bibitem[\protect\citeauthoryear{Sinha and Swearingen}{Sinha and
  Swearingen}{2002}]%
        {sinha2002role}
\bibfield{author}{\bibinfo{person}{Rashmi Sinha} {and} \bibinfo{person}{Kirsten
  Swearingen}.} \bibinfo{year}{2002}\natexlab{}.
\newblock \showarticletitle{The role of transparency in recommender systems}.
  In \bibinfo{booktitle}{\emph{CHI'02 extended abstracts on Human factors in
  computing systems}}. ACM, \bibinfo{pages}{830--831}.
\newblock


\bibitem[\protect\citeauthoryear{Sun, Wu, Zhang, Fu, Hong, and Wang}{Sun
  et~al\mbox{.}}{2020}]%
        {sun2020dual}
\bibfield{author}{\bibinfo{person}{Peijie Sun}, \bibinfo{person}{Le Wu},
  \bibinfo{person}{Kun Zhang}, \bibinfo{person}{Yanjie Fu},
  \bibinfo{person}{Richang Hong}, {and} \bibinfo{person}{Meng Wang}.}
  \bibinfo{year}{2020}\natexlab{}.
\newblock \showarticletitle{Dual Learning for Explainable Recommendation:
  Towards Unifying User Preference Prediction and Review Generation}. In
  \bibinfo{booktitle}{\emph{Proceedings of The Web Conference 2020}}.
  \bibinfo{pages}{837--847}.
\newblock


\bibitem[\protect\citeauthoryear{Sutskever, Vinyals, and Le}{Sutskever
  et~al\mbox{.}}{2014}]%
        {sutskever2014sequence}
\bibfield{author}{\bibinfo{person}{Ilya Sutskever}, \bibinfo{person}{Oriol
  Vinyals}, {and} \bibinfo{person}{Quoc~V Le}.}
  \bibinfo{year}{2014}\natexlab{}.
\newblock \showarticletitle{Sequence to sequence learning with neural
  networks}. In \bibinfo{booktitle}{\emph{Advances in neural information
  processing systems}}. \bibinfo{pages}{3104--3112}.
\newblock


\bibitem[\protect\citeauthoryear{Tao, Jia, Wang, and Wang}{Tao
  et~al\mbox{.}}{2019}]%
        {tao2019the}
\bibfield{author}{\bibinfo{person}{Yiyi Tao}, \bibinfo{person}{Yiling Jia},
  \bibinfo{person}{Nan Wang}, {and} \bibinfo{person}{Hongning Wang}.}
  \bibinfo{year}{2019}\natexlab{}.
\newblock \showarticletitle{The FacT: Taming Latent Factor Models for
  Explainability with Factorization Trees}. In
  \bibinfo{booktitle}{\emph{Proceedings of the 42Nd International ACM SIGIR
  Conference on Research and Development in Information Retrieval}}.
  \bibinfo{publisher}{ACM}, \bibinfo{address}{New York, NY, USA},
  \bibinfo{pages}{295--304}.
\newblock


\bibitem[\protect\citeauthoryear{Truong and Lauw}{Truong and Lauw}{2019}]%
        {truong2019multimodal}
\bibfield{author}{\bibinfo{person}{Quoc-Tuan Truong} {and}
  \bibinfo{person}{Hady Lauw}.} \bibinfo{year}{2019}\natexlab{}.
\newblock \showarticletitle{Multimodal Review Generation for Recommender
  Systems}. In \bibinfo{booktitle}{\emph{The World Wide Web Conference}}.
  \bibinfo{pages}{1864--1874}.
\newblock


\bibitem[\protect\citeauthoryear{Vezhnevets, Osindero, Schaul, Heess,
  Jaderberg, Silver, and Kavukcuoglu}{Vezhnevets et~al\mbox{.}}{2017}]%
        {vezhnevets2017feudal}
\bibfield{author}{\bibinfo{person}{Alexander~Sasha Vezhnevets},
  \bibinfo{person}{Simon Osindero}, \bibinfo{person}{Tom Schaul},
  \bibinfo{person}{Nicolas Heess}, \bibinfo{person}{Max Jaderberg},
  \bibinfo{person}{David Silver}, {and} \bibinfo{person}{Koray Kavukcuoglu}.}
  \bibinfo{year}{2017}\natexlab{}.
\newblock \showarticletitle{Feudal networks for hierarchical reinforcement
  learning}. In \bibinfo{booktitle}{\emph{International Conference on Machine
  Learning}}. PMLR, \bibinfo{pages}{3540--3549}.
\newblock


\bibitem[\protect\citeauthoryear{Wang, Lu, and Zhai}{Wang
  et~al\mbox{.}}{2010}]%
        {wang2010latent}
\bibfield{author}{\bibinfo{person}{Hongning Wang}, \bibinfo{person}{Yue Lu},
  {and} \bibinfo{person}{Chengxiang Zhai}.} \bibinfo{year}{2010}\natexlab{}.
\newblock \showarticletitle{Latent aspect rating analysis on review text data:
  a rating regression approach}. In \bibinfo{booktitle}{\emph{Proceedings of
  the 16th ACM SIGKDD international conference on Knowledge discovery and data
  mining}}. \bibinfo{pages}{783--792}.
\newblock


\bibitem[\protect\citeauthoryear{Wang, Wang, Jia, and Yin}{Wang
  et~al\mbox{.}}{2018b}]%
        {wang2018explainable}
\bibfield{author}{\bibinfo{person}{Nan Wang}, \bibinfo{person}{Hongning Wang},
  \bibinfo{person}{Yiling Jia}, {and} \bibinfo{person}{Yue Yin}.}
  \bibinfo{year}{2018}\natexlab{b}.
\newblock \showarticletitle{Explainable recommendation via multi-task learning
  in opinionated text data}. In \bibinfo{booktitle}{\emph{The 41st
  International ACM SIGIR Conference on Research \& Development in Information
  Retrieval}}. \bibinfo{pages}{165--174}.
\newblock


\bibitem[\protect\citeauthoryear{Wang, Chen, Yang, Wu, Wu, and Xie}{Wang
  et~al\mbox{.}}{2018a}]%
        {wang2018reinforcement}
\bibfield{author}{\bibinfo{person}{Xiting Wang}, \bibinfo{person}{Yiru Chen},
  \bibinfo{person}{Jie Yang}, \bibinfo{person}{Le Wu},
  \bibinfo{person}{Zhengtao Wu}, {and} \bibinfo{person}{Xing Xie}.}
  \bibinfo{year}{2018}\natexlab{a}.
\newblock \showarticletitle{A reinforcement learning framework for explainable
  recommendation}. In \bibinfo{booktitle}{\emph{2018 IEEE International
  Conference on Data Mining (ICDM)}}. IEEE, \bibinfo{pages}{587--596}.
\newblock


\bibitem[\protect\citeauthoryear{Welleck, Kulikov, Roller, Dinan, Cho, and
  Weston}{Welleck et~al\mbox{.}}{2019}]%
        {welleck2019neural}
\bibfield{author}{\bibinfo{person}{Sean Welleck}, \bibinfo{person}{Ilia
  Kulikov}, \bibinfo{person}{Stephen Roller}, \bibinfo{person}{Emily Dinan},
  \bibinfo{person}{Kyunghyun Cho}, {and} \bibinfo{person}{Jason Weston}.}
  \bibinfo{year}{2019}\natexlab{}.
\newblock \showarticletitle{Neural text generation with unlikelihood training}.
\newblock \bibinfo{journal}{\emph{arXiv preprint arXiv:1908.04319}}
  (\bibinfo{year}{2019}).
\newblock


\bibitem[\protect\citeauthoryear{Weston, Dinan, and Miller}{Weston
  et~al\mbox{.}}{2018}]%
        {weston2018retrieve}
\bibfield{author}{\bibinfo{person}{Jason Weston}, \bibinfo{person}{Emily
  Dinan}, {and} \bibinfo{person}{Alexander~H Miller}.}
  \bibinfo{year}{2018}\natexlab{}.
\newblock \showarticletitle{Retrieve and refine: Improved sequence generation
  models for dialogue}.
\newblock \bibinfo{journal}{\emph{arXiv preprint arXiv:1808.04776}}
  (\bibinfo{year}{2018}).
\newblock


\bibitem[\protect\citeauthoryear{Wu, Cai, and Wang}{Wu et~al\mbox{.}}{2020}]%
        {wu2020deja}
\bibfield{author}{\bibinfo{person}{Jibang Wu}, \bibinfo{person}{Renqin Cai},
  {and} \bibinfo{person}{Hongning Wang}.} \bibinfo{year}{2020}\natexlab{}.
\newblock \showarticletitle{D{\'e}j{\`a} vu: A contextualized temporal
  attention mechanism for sequential recommendation}. In
  \bibinfo{booktitle}{\emph{Proceedings of The Web Conference 2020}}.
  \bibinfo{pages}{2199--2209}.
\newblock


\bibitem[\protect\citeauthoryear{Xian, Fu, Muthukrishnan, De~Melo, and
  Zhang}{Xian et~al\mbox{.}}{2019}]%
        {xian2019reinforcement}
\bibfield{author}{\bibinfo{person}{Yikun Xian}, \bibinfo{person}{Zuohui Fu},
  \bibinfo{person}{Shan Muthukrishnan}, \bibinfo{person}{Gerard De~Melo}, {and}
  \bibinfo{person}{Yongfeng Zhang}.} \bibinfo{year}{2019}\natexlab{}.
\newblock \showarticletitle{Reinforcement knowledge graph reasoning for
  explainable recommendation}. In \bibinfo{booktitle}{\emph{Proceedings of the
  42nd international ACM SIGIR conference on research and development in
  information retrieval}}. \bibinfo{pages}{285--294}.
\newblock


\bibitem[\protect\citeauthoryear{Xian, Zhao, Li, Chan, Kan, Ma, Dong,
  Faloutsos, Karypis, Muthukrishnan, et~al\mbox{.}}{Xian et~al\mbox{.}}{2021}]%
        {xian2021ex3}
\bibfield{author}{\bibinfo{person}{Yikun Xian}, \bibinfo{person}{Tong Zhao},
  \bibinfo{person}{Jin Li}, \bibinfo{person}{Jim Chan}, \bibinfo{person}{Andrey
  Kan}, \bibinfo{person}{Jun Ma}, \bibinfo{person}{Xin~Luna Dong},
  \bibinfo{person}{Christos Faloutsos}, \bibinfo{person}{George Karypis},
  \bibinfo{person}{Shan Muthukrishnan}, {et~al\mbox{.}}}
  \bibinfo{year}{2021}\natexlab{}.
\newblock \showarticletitle{EX3: Explainable Attribute-aware Item-set
  Recommendations}. In \bibinfo{booktitle}{\emph{Fifteenth ACM Conference on
  Recommender Systems}}. \bibinfo{pages}{484--494}.
\newblock


\bibitem[\protect\citeauthoryear{Yang, Wang, Deng, and Wang}{Yang
  et~al\mbox{.}}{2021}]%
        {yang2021explanation}
\bibfield{author}{\bibinfo{person}{Aobo Yang}, \bibinfo{person}{Nan Wang},
  \bibinfo{person}{Hongbo Deng}, {and} \bibinfo{person}{Hongning Wang}.}
  \bibinfo{year}{2021}\natexlab{}.
\newblock \showarticletitle{Explanation as a Defense of Recommendation}. In
  \bibinfo{booktitle}{\emph{Proceedings of the 14th ACM International
  Conference on Web Search and Data Mining}}. \bibinfo{pages}{1029--1037}.
\newblock


\bibitem[\protect\citeauthoryear{Zhang and Chen}{Zhang and Chen}{2020}]%
        {zhang2018explainable}
\bibfield{author}{\bibinfo{person}{Yongfeng Zhang} {and} \bibinfo{person}{Xu
  Chen}.} \bibinfo{year}{2020}\natexlab{}.
\newblock \showarticletitle{Explainable recommendation: A survey and new
  perspectives}.
\newblock \bibinfo{journal}{\emph{Foundations and Trends{\textregistered} in
  Information Retrieval}} \bibinfo{volume}{14}, \bibinfo{number}{1}
  (\bibinfo{year}{2020}), \bibinfo{pages}{1--101}.
\newblock


\bibitem[\protect\citeauthoryear{Zhang, Lai, Zhang, Zhang, Liu, and Ma}{Zhang
  et~al\mbox{.}}{2014}]%
        {zhang2014explicit}
\bibfield{author}{\bibinfo{person}{Yongfeng Zhang}, \bibinfo{person}{Guokun
  Lai}, \bibinfo{person}{Min Zhang}, \bibinfo{person}{Yi Zhang},
  \bibinfo{person}{Yiqun Liu}, {and} \bibinfo{person}{Shaoping Ma}.}
  \bibinfo{year}{2014}\natexlab{}.
\newblock \showarticletitle{Explicit factor models for explainable
  recommendation based on phrase-level sentiment analysis}. In
  \bibinfo{booktitle}{\emph{Proceedings of the 37th international ACM SIGIR
  conference on Research \& development in information retrieval}}.
  \bibinfo{pages}{83--92}.
\newblock


\end{thebibliography}

%%
%% If your work has an appendix, this is the place to put it.

\appendix
% \clearpage
\renewcommand{\arraystretch}{1.0}
\begin{table*}[t]
    \caption{Case study of the explanations generated by different models.}
    \vspace{-2mm}
    \setlength\tabcolsep{4pt}
    \label{tab:sample}
    \begin{tabular}{|c|l|l|}
        \hline
        Model & Sample 1  & Sample 2 \\
        %  & Hantverksbryggeriet Baronen & Key West Southernmost Wheat \\
        \hline
        Human & aroma of caramel, cherry, raisins, and florals. & pours clear yellow body with a small white head. \\
        NARRE & the finish is dry and ashy. & not bad, if one is looking for a refreshing, light wheat beer. \\
        NRT & flavor of chocolate, roasted malt, and light smoke. & the beer is a hazy yellow-orange color. \\
        SAER & aroma of caramel, caramel, and citrus. & medium body, watery texture, and carbonation. \\
        \model{}-Ext & sweet aroma with toasted malt, caramel and alcohol notes.& pours a hazy golden with a small white head. \\

        \model{} & aroma of caramel, malt, and alcohol. & pours a hazy yellow body with a small white head.\\
        \hline
    \end{tabular}
    % \vspace{-2mm}
\end{table*}

% \section{Supplementary}

\section{Model Implementation Details}
In the section, we will share our technical choices of some important components and hyper-parameter values in \model{}.

\model{}'s extractor adopts a single text encoder to obtain universal sentence representations for reviews from both user and item profile. The text encoder's architecture follows the self-attentive network presented in \cite{conneau2017supervised}, where the attention mechanism aggregates the hidden states of a bi-directional GRU. The GRU is of a single layer with hidden state size of 300. For the input word embeddings, we bootstrap their initial values with GloVe 6B of 300 dimensions \cite{pennington2014glove} and allow them to be further updated during training.   

As we discussed before, our proposed solution is able to handle both continuous and discrete ratings, but in this work, we assume recommendation ratings are discrete. Therefore, our implementation applies a rating embeddings to map the one-hot vector of rating difference into its latent representation. These rating embeddings are randomly initialized and learned through the back-propagation during the training process. We set the embedding size to 16. For the following latent space transformation $f(x^u_i,\Delta r^u_i) \to h_i$, we use a 2-layer MLP with Tanh as the activation function, whose intermediate and final output sizes are both 300.

The final text decoder inside \model{}'s refiner is another single layer GRU with hidden size of 300. The text encoder from the extractor and this decoder together actually forms a sequence-to-sequence model \cite{sutskever2014sequence} whose input is the chosen prototype sentence. So our extract-and-refine process can be viewed as multiple sequence encoders run in parallel, while only one of them can connect to the sequence decoder. Additionally, the text decoder also adopts the attention layer proposed in \cite{luong2015effective}. We find that paying attention to the extracted prototype during the refining process is beneficial to the clipped recall defined in Eq \eqref{eq_clipped_recall}.

Since this work focuses on studying the problem of comparative explanation, the above techniques are enough for us to demonstrate the effectiveness of our proposed solution. The architecture of \model{} itself does not hold any assumptions about the implementation of the discussed sub-components. They can be replaced with other state-of-the-art models to further boost the performance.

\section{Model Training Details}

In the section, we will discuss the techniques we used to train \model{} and the corresponding hyper-parameters.

To train the extractor, we need to batch multiple user profiles and item profile respectively. However, the sizes of the profiles vary a lot. When we batch the profiles with all their reviews, the batch will end with many paddings to ensure every profile within the batch has the same size. These paddings waste lots of computing resources and heavily slow down the training process. So instead of using all the reviews, we define a max limit. For a profile larger than the limit, we will randomly sample a subset based on the limit. Larger limit usually leads to better training results since the model have more references and candidates to leverage. We set the limit to 10 for both user and item based on our computing capacity and we also found the improvement beyond it is marginal.

The value of $\kappa$ from Eq \eqref{eq_extract_prob} is critical to the training since it balances the exploration and exploitation in the policy gradient. Smaller values flatten the extraction distribution and hence force the model to explore more extraction candidates, but this tends to delay the convergence and cause very unstable results. On the other hand, larger values concentrate the distribution and reduce the search space, but then the model may miss more appropriate candidates and lose the meaning of the joint training. Based on our tests, we find 3 is the most balanced value.

IDF-BLEU is the main reward in the policy gradient training, but we used unconventional n-gram weights. Following the original design of BLEU, IDF-BLEU keeps the individual weight for each type of n-gram precision, i.e., $w_n$ in Eq \eqref{eq_BLEU}. The weights we used for unigram to 4-gram are \{0.8, 0.2, 0, 0\}. Usually, the weights are equally distributed among all the available n-grams. For example, BLEU-2 has the weight of 0.5 for both unigram and bigram; similarly, IDF-BLEU-4 applies a unified weight of 0.25 from unigram to 4-gram. However, the precision of different n-grams are not always compatible with each other as objectives and the model has to make trade-offs. For example, we found using IDF-BLEU-4 as reward sacrifice unigram and bigram precision in exchange for 4-gram precision. As a result, it only slightly benefits the IDF-BLEU-4 in evaluation but leads IDF-BLEU-\{1,2\} to decline. Therefore, we decided this customized weights to only focus on unigram and bigram. There are two reasons. First, correctly generating trigrams and 4-grams in explanations is quite difficult so such overlaps will not be very frequent anyway. Instead of betting for some dull 4-grams, it is more valuable to cover the interested features which are often just unigrams. Second, higher precision of unigram and bigram still contribute to IDF-BLEU-4 in evaluation according to its definition. But they may not always compensate the negligence of other n-grams. This explains why \model{} has a less competitive IDF-BLEU-4 in TripAdvisor meanwhile leading the rest in Table \ref{tab:exp_eval}.

% This is because correctly generating trigrams and 4-grams is quite difficult in explanation generation so these overlaps will not be not very frequent. U

\section{Case Study}
% \subsubsection{Case Study.}
% \noindent\textbf{Case Study.}
Groups of example explanations generated by \model{} and other baselines are shown in Table \ref{tab:sample}. 
%We also included results from \model{}-Ext to demonstrate how the extract-and-refine flow works. 
The ground-truth explanations are given for reference denoted as \emph{Human}. 
Comparing NARRE and \model{}-Ext shows the value of modeling comparativeness in users provided historical content. Sentences extracted by \model{}-Ext are much closer to the ground-truth than NARRE's. 
Comparing \model{}-Ext and \model{} shows the effectiveness our rewriting module in improving the explanation quality, especially in writing style and wording. For example, in Sample 1, the extracted explanation correctly covers the attribute ``aroma'' and ``caramel'', but its sentence structure is different from the ground-truth's. The refined explanation keeps the two correct attributes and improves the sentence structure. In Sample 2, while the extractor  picks a sentence almost the same as the ground-truth, the refiner further changes the word ``golden'' to ``yellow'', which better reflects the user's preference in wording. However, both samples also suggest our refiner can be further improved in personalized feature revision. For example, in Sample 2, the end explanation inherits "hazy" from the extracted prototype while the item looks "clear" instead for the target user. Same for "malt" and "alcohol" vs., "cherry" and "raisins" in Sample 1. Obviously, these features are subjective and the target user hold a different opinion from the author of the extracted sentence. It would be a promising future direction to better personalize subjective features while still maintain the relevance and faithfulness to other objective facts given by the extraction.

\end{document}